\documentclass[]{article}

\usepackage{chemformula} 
\usepackage[T1]{fontenc} 
\usepackage{graphicx, xcolor} 
\usepackage{amsmath, gensymb} 
\usepackage{hhline}
\usepackage[caption=false]{subfig}
\usepackage{siunitx}
\usepackage{miller}
\usepackage{xr}
\usepackage{upgreek}
\usepackage{authblk}
\usepackage{fullpage}
\usepackage{hyperref}
\graphicspath{{Figures//}}

\makeatletter
\newcommand*{\addFileDependency}[1]{
  \typeout{(#1)}
  \@addtofilelist{#1}
  \IfFileExists{#1}{}{\typeout{No file #1.}}
}
\makeatother

\newcommand*{\myexternaldocument}[1]{%
    \externaldocument{#1}%
    \addFileDependency{#1.tex}%
    \addFileDependency{#1.aux}%
}

\myexternaldocument{Supplementary}

\newcommand{\AlOx}{Al$_2$O$_3$}
\newcommand{\PrUnits}{$\upmu$C/cm$^2$}
\newcommand{\HZO}{Hf$_{0.5}$Zr$_{0.5}$O$_2$}
\newcommand{\HfO}{HfO$_2$}

\title{Towards Non-volatile Spin Orbit Devices – Deposition of Ferroelectric Hafnia on Monolayer Graphene/Co/HM Stacks} 
\author[1]{Suzanne Lancaster*}
\author[2]{Iciar Arnay}
\author[2,3]{Ruben Guerrero}
\author[2]{Adrian Gud\'{i}n}
\author[4]{Alejandra Guedeja-Marr\'{o}n}
\author[5]{Jose Manuel Diez}
\author[1]{Jan G\"{a}rtner}
\author[2]{Alberto Anad\'{o}n}
\author[4]{Maria Varela}
\author[5]{Julio Camarero}
\author[1,6]{Thomas Mikolajick}
\author[2]{Paolo Perna} 
\author[1]{Stefan Slesazeck}

\affil[1]{NaMLab gGmbH, N\"{o}thnitzer Str. 64a, 01187 Dresden, Germany}
\affil[2]{IMDEA Nanociencia, c/ Faraday 9, 28049 Madrid, Spain}
\affil[3]{Current address: University of Castilla-La Mancha, 13001 Ciudad Real, Spain}
\affil[4]{Departamento de F\'{i}sica de Materiales and Instituto Pluridisciplinar, Universidad Complutense de Madrid, Ciudad Universitaria, 28040 Madrid, Spain}
\affil[5]{Departamento de F\'{i}sica de la Materia Condensada \& Departamento de F\'{i}sica Aplicada \& Instituto Nicol\'{a}s Cabrera, Universidad Aut\'{o}noma de Madrid, 28049 Madrid, Spain}
\affil[6]{Institute of Semiconductors and Microsystems, Technische Universit\"{a}t Dresden, N\"{o}thnitzer Str 64., 01187 Dresden, Germany}
\affil[*]{Corresponding author. Email: suzanne.lancaster@namlab.com}

\begin{document}
\date{}
\maketitle 

\begin{abstract}
  While technologically challenging, the integration of ferroelectric thin films with graphene spintronics potentially allows the realization of highly efficient, electrically tuneable, non-volatile memories through control of the interfacial spin-orbit driven interaction occuring at graphene/Co interfaces deposited on heavy metal supports.

Here, the integration of ferroelectric \HZO{} on graphene/Co/heavy metal epitaxial stacks is investigated via the implementation of several nucleation methods in atomic layer deposition. By employing in-situ \AlOx{} as a nucleation layer sandwiched between \HZO{} and graphene, the \HZO{} demonstrates a remanent polarization (\textit{2Pr}) of 19.2 \PrUnits. Using an ex-situ, naturally oxidized sputtered Ta layer for nucleation, \textit{2Pr} could be controlled via the interlayer thickness, reaching maximum values of 28 \PrUnits{} with low coercive fields. 

Magnetic hysteresis measurements taken before and after atomic layer deposition show strong perpendicular magnetic anisotropy, with minimal deviations in the magnetization reversal pathways due to the \HZO{} deposition process, thus pointing to a good preservation of the magnetic stack including single-layer graphene. X-ray diffraction measurements further confirm that the high-quality interfaces demonstrated in the stack remain unperturbed by the ferroelectric deposition and anneal.

The proposed graphene-based ferroelectric/magnetic structures offer the strong advantages of ferroelectricity and ferromagnetism at room temperature, enabling the development of novel magneto-electric and non-volatile in-memory spin-orbit logic architectures with low power switching. \footnote{This article was accepted at ACS Applied Materials \& Interfaces and can be found at \href{https://doi.org/10.1021/acsami.2c22205}{https://doi.org/10.1021/acsami.2c22205}. This article may be downloaded for personal use only. Any other use requires prior permission of the author and ACS Publications.}
\end{abstract}

\section{Introduction}
Graphene, besides being considered as an interesting candidate for traditional electronics \cite{novoselov2012roadmap}, has recently found applications in spintronics \cite{roche2015graphene,piquemal2020spin}, offering long spin lifetimes and long spin propagation lengths \cite{han2014graphene}. While graphene exhibits negligible intrinsic spin-orbit coupling (SOC), it can be amplified by proximity effects \cite{avsar2014spin, anadon2021engineering}, for example when integrated with a heavier atom \cite{marchenko2012giant,calleja2015spatial,klimovskikh2017spin}. In addition, it has been shown that the graphene-ferromagnet interface enhances interfacial perpendicular magnetic anisotropy (PMA) \cite{rougemaille2012perpendicular, ajejas2018unraveling} in Co/heavy metal (HM) stacks. This improvement is due to the enhancement of orbital magnetic moment anisotropy promoted by the presence of Gr through the hybridization between the $\pi$-band of Gr with the 3d-orbital of Co \cite{ajejas2018unraveling,blanco2021large}, extending the Co thickness range for exhibiting PMA up to 4 nm. Further, such Gr/Co interfaces present sizeable Dzyaloshinskii-Moriya interaction (DMI) \cite{ajejas2018unraveling, yang2018significant}, which is critical for stabilizing skyrmionic spin textures \cite{fert2013skyrmions} and is particularly interesting in the case of graphene \cite{olleros2020intrinsic}, as the interfacial magnetic interactions may be efficiently electrically tuned.

At the same time, the discovery of ferroelectricity in doped \HfO{} thin films \cite{boscke2011ferroelectricity} has opened up a new avenue for ferroelectric research, with many benefits compared to classic perovskite materials, particularly for process integration \cite{park2018review}. \HfO{} films deposited via atomic layer deposition (ALD) provide good process compatibility with chemical vapor deposition (CVD), which has been demonstrated as a promising technique for the scalable growth of single-layer graphene \cite{reina2009large}. The high quality and versatility of \HfO{} films has already been demonstrated in non-volatile memory devices such as capacitor-based random access memories, ferroelectric field-effect transistors (FETs) and ferroelectric tunnel junctions \cite{martin2013downscaling, fujii2016first}.

 Previously, graphene has been explored as a channel for ferroelectric FETs via transfer methods \cite{ahn2018carbon, dragoman2020memtransistors}. The integration of ferroelectrics on graphene and metal stacks with atomically sharp interfaces allows the potential for new device geometries exploiting the interfacial Rashba SOC which has been demonstrated in these systems \cite{anadon2021engineering, marchenko2012giant, cano2022rashba}. In fact, in Rashba systems, the ferroelectric control of spin-charge conversion at 4\,K was recently demonstrated in spin-orbit logic devices \cite{noel2020non} similar to a magneto-electrical spin-orbit logic architecture proposed by Intel \cite{manipatruni2019scalable}. The materials demonstrated here would be ideal candidates for spin-orbit logic given that large spin-charge conversion has been shown in graphene \cite{anadon2021engineering} and \HZO{} is a room-temperature ferroelectric.
 
 In addition, the electrically tunable DMI at the graphene-Co interface \cite{ajejas2018unraveling} makes it suitable for devices based on magnetic skyrmions. Skyrmion electronics are highly anticipated due to the topologically robust nature and low driving currents of magnetic skyrmions \cite{fert2013skyrmions}, as well as their small size which can be controlled via the DMI \cite{buttner2018theory}. Thus, spintronic and skyrmion-based devices offer promise for fast, power efficient, non-volatile in-memory logic operation which is imperative for the next generation of computing architectures \cite{covi2022challenges, covi2021ferroelectric}. 

To-date, the integration of ferroelectrics on graphene mainly focuses on spin-coated polymers \cite{zheng2010graphene} or the exfoliation of graphene onto ferroelectric substrates \cite{jie2014graphene}, either perovskites \cite{lee2013flexible} or, most recently, ferroelectric HZO thin films \cite{dragoman2018wafer}. For the chemical vapour deposition (CVD) of graphene, where large-scale, single-layer graphene (SLG) can be reliably deposited on various metal substrates \cite{reina2009large}, subsequent transfer of the film can lead to a deterioration in material quality \cite{kang2012graphene}, while also losing any scaling benefits and the precise control over interface properties. Moreover, voltage-control of magnetic anisotropy has been demonstrated at CoPt/Al:HfO2 interfaces \cite{vermeulen2019ferroelectric}, but disordered, intermixed Co-Pt interfaces may reduce the DMI strength \cite{zimmermann2018dzyaloshinskii}.  In this paper, we demonstrate the direct integration of a \HZO{} (HZO) ferroelectric thin film onto CVD-grown graphene on Co/Pt bilayers via atomic layer deposition (ALD). The method can be applied without any intermixing of the Co/Pt or damage to the graphene layer, evidenced by magnetic hysteresis measurements. This finding opens up new possibilities for the engineering of layer stacks containing graphene, while maintaining the advantages of large (mm)-scale integration and a high material quality.

\section{In-situ methods for nucleation on the graphene surface}
\subsection{Film deposition and characterization}
Atomic layer deposition of dielectric materials on graphene (Gr) and other 2D materials depends sensitively on surface functionalization, precursor and oxidant choice, and temperature, as summarized in \textit{Kim et al.} \cite{kim2017atomic}. In particular, chemisorption on the graphene surface, such as that required for the ALD process, is generally preferred at defect sites \cite{qi2012density} and may be difficult to engineer on high-quality graphene surfaces.

The HZO films presented here were deposited with HyALD and ZyALD as Hf and Zr precursor, respectively. These precursors cannot be used with H$_2$O as an oxidant, which has been shown to have good wetting of Gr at lower temperatures, so that physisorption can be used in the absence of chemisorption to promote film growth \cite{zhang2014direct}. Of the remaining oxidants, O$_3$, while demonstrated to functionalize the Gr surface at room temperature, is damaging to SLG at the elevated temperatures required for ferroelectric deposition \cite{jandhyala2012atomic}. Furthermore, the choice of oxidant has been shown to limit the temperature range in which the ferroelectric orthorhombic phase can be formed\cite{alcala2020influence}, and for applications in spin-orbitronics which exploit interfacial effects, lower temperatures are needed to prohibit intermixing of Co and Pt \cite{ajejas2019thermally}. Therefore, a method was developed using a remote O$_2$ plasma as an oxidant, with the plasma source located around 30 cm from the substrate. Under these conditions, the damage to SLG should be minimized \cite{tang2015damage} and O$_2$ plasma has been shown to produce high-quality ALD films on graphene \cite{canto2021plasma}.

\begin{figure}[ht]
\centering
		\includegraphics[height=9cm]{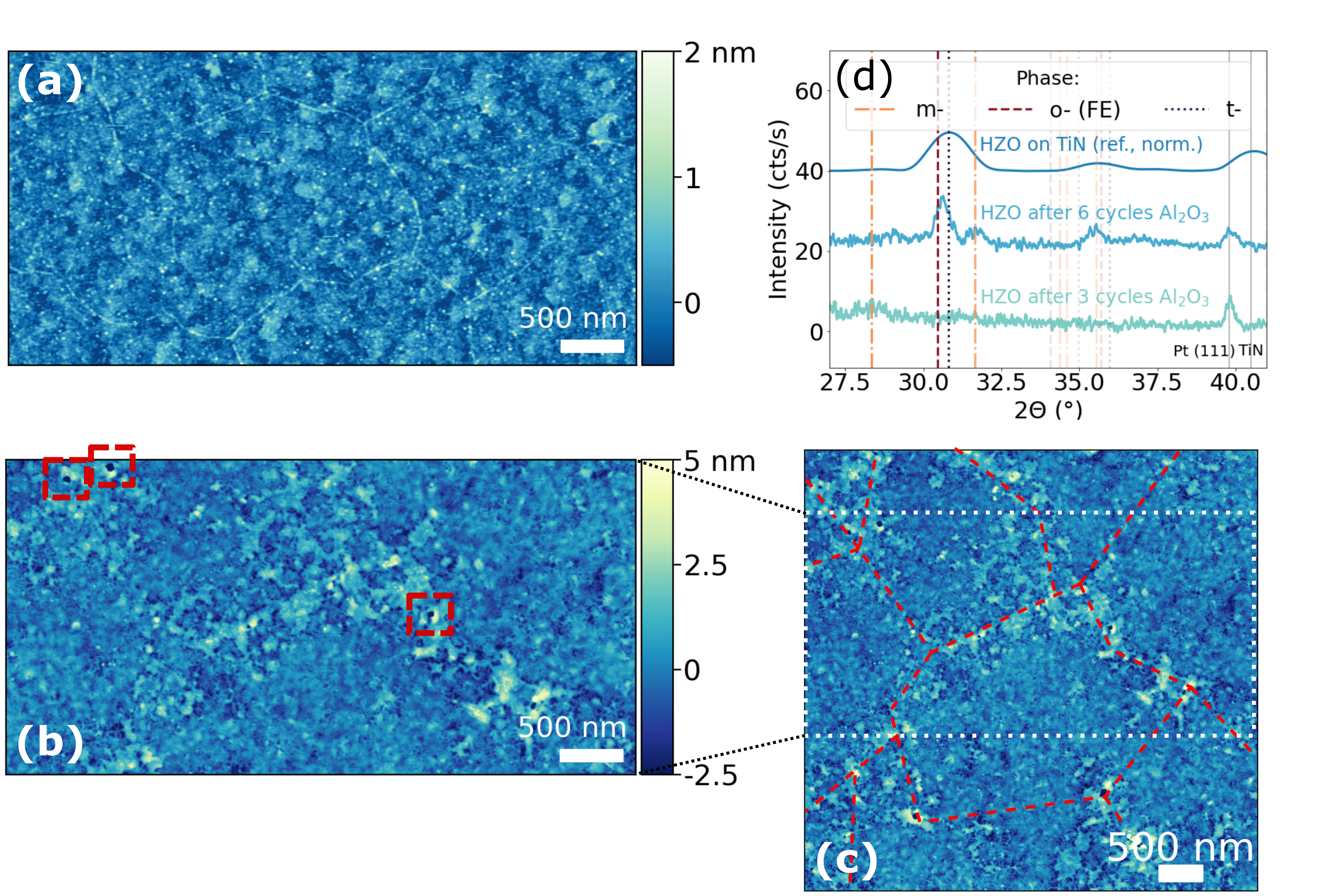}

	\captionsetup{font=footnotesize}
	\caption{AFM study of in-situ nucleation steps: (a) after three extended cycles HyALD with O$_2$ as oxidant and 110 cycles of HZO deposition; (b) after 3 cycles TMA with H$_2$O as oxidant; (c) full AFM map (the white dotted lines indicate the reduced area shown in (b)) with step edges marked with red dashed lines. (d) GIXRD scans of HZO films deposited after 3 cycles (green) and 6 cycles (blue) of \AlOx{} deposition. 
	}
\label{fig1_AFM}
\end{figure}

Figure \ref{fig1_AFM} shows the results of an atomic force microscopy (AFM, measured in tapping mode on a Bruker Dimension XR microscope) and grazing incidence X-ray diffraction (GIXRD, measured with a Cu K-$\alpha$1 X-ray source on a Bruker D8 Discover XRD system) study on in-situ nucleation methods on the Gr surface of Gr/Pt(111) stacks, with a Pt thickness of 30 nm. First, HZO deposition directly on the Gr surface was investigated, with no additional interlayer. Films were deposited in an Oxford OpAL ALD system, with alternating ALD cycles of HyALD/O$_2$ plasma and ZyALD/O$_2$ plasma. For the first three deposition cycles, HfO$_2$ only was deposited and the HyALD precursor residence time was increased to promote chemisorption, followed by deposition of HZO and finally \AlOx{} using TMA/H$_2$O (full details are given in the supplementary materials). A subsequent AFM map was recorded and is shown in figure \ref{fig1_AFM}(a). While no conformal film was deposited, decorated lines appear, indicating enhanced reactivity at some sites on the Gr surface. The size of these features, (1-2\,$\upmu$m), is consistent with step edges previously observed in scanning tunneling microscopy (STM) experiments on the same stacks \cite{ajejas2019thermally}. Depending on the ALD precursor used, the energy barrier for chemisorption is increased \cite{kim2014selective} and thus nucleation is favoured only at more reactive sites, such as these step edges. As shown in the supplementary materials (see supplementary information), the average difference in height either side of the line features is $0.242 \pm 0.039$\,nm. This corresponds well with a single step edge in Pt(111), 0.238\,nm \cite{itaya1990nsitu}.

Due to the poor nucleation on the pristine graphene surface, subsequent experiments focused on adding an interlayer between the Gr and HZO. Employing a dielectric interlayer would still allow for electric field control over the Gr/Co interface, as has been widely researched in the context of FeFET devices.\cite{ni2018critical} At the same time, spin-orbit effects can be tuned both through a dielectric layer \cite{srivastava2018large} or even metallic interlayer \cite{anadon2020spin} for controlling the DMI.

To investigate nucleation performed in-situ by ALD, a seed layer of Al$_2$O$_3$ was deposited, followed by 11\,nm HZO and a capping layer for crystallization. For the seed layer only, a lower ALD temperature of 150$\degree$C was applied. When depositing Al$_2$O$_3$ with a trimethylaluminum (TMA) precursor,  H$_2$O can be employed as an oxidant and deposition can occur at a reduced temperature \cite{zhang2014direct}. In these samples, the metallic layer is expected to play a role in the ALD process by further increasing wetting of the surface by H$_2$O \cite{dlubak2012substrate}. The lower temperature should help to avoid damage to the Gr as well as prevent possible intermixing of Co/HM \cite{ajejas2019thermally}.

Figure \ref{fig1_AFM}(b) shows an AFM map of the sample surface after 3 deposition cycles of the Al$_2$O$_3$ interlayer. Enhanced nucleation can still be clearly observed at step edges in Pt(111), indicated by the red dashed lines on the larger-scale image (figure \ref{fig1_AFM}(c)). Nonetheless, the improved surface coverage is visible in AFM and is apparent in the increased average roughness of 0.69\,nm, pointing to a more efficient nucleation even after very few ALD cycles. An additional feature observed in AFM images is the existence of large structural defects, highlighted in red boxes in figure \ref{mag_hyst}, which are seen to propagate through the HZO films (see supplementary information) and may reduce the breakdown field of the \AlOx{} layer. It should be noted that the slight deviations in height introduced by the selective nucleation on \AlOx{} appear to reduce as the film thickness increases, and are absent in scanning electron microscopy (SEM) and AFM images after HZO deposition, as shown in supplementary.

Films of $\sim$ 15\,nm HZO were deposited after 3 or 6 nucleation cycles of Al$_2$O$_3$ (see supplementary materials for full deposition details). In order to assess the crystallization of the deposited HZO, GIXRD measurements were first performed to identify whether the HZO crystallizes and in which phases. Figure \ref{fig1_AFM}(d) shows 2$\theta$ scans for the samples with 3 nucleation cycles (green) and 6 nucleation cycles (blue), measured after annealing for 5\,s at 600$\degree$C. Peaks first appear in the GIXRD spectrum for 6 nucleation cycles, indicating polycrystalline HZO, indicating that a minimum \AlOx{} thickness is needed as a nucleation layer. This is likely due to non-uniform surface coverage at smaller thicknesses. Comparison to the powder diffraction database (PDF) entries for different crystal phases of HfO$_2$ (dashed lines in figure \ref{fig1_AFM}(d)) confirms that the HZO is composed of a mixture of the non-ferroelectric monoclinic (m-) phase, ferroelectric orthorhombic (o-) phase and antiferroelectric tetragonal (t-) phase. The main peak in the GIXRD scan is at around 31\degree, which is consistent with PDF values for a mixed o-/t-phase, and the peak position is comparable to a reference spectrum measured on HZO with TiN electrodes. On the other hand, a significant m-phase portion is visible at 28\degree{} and 32\degree{}, larger than in the reference sample. Peak fitting is presented and further discussed in the supplementary materials (see supplementary information). 

\subsection{Electrical characterization of in-situ nucleated films}

\begin{figure}[!bht]
\begin{center}
		\includegraphics[width=15cm]{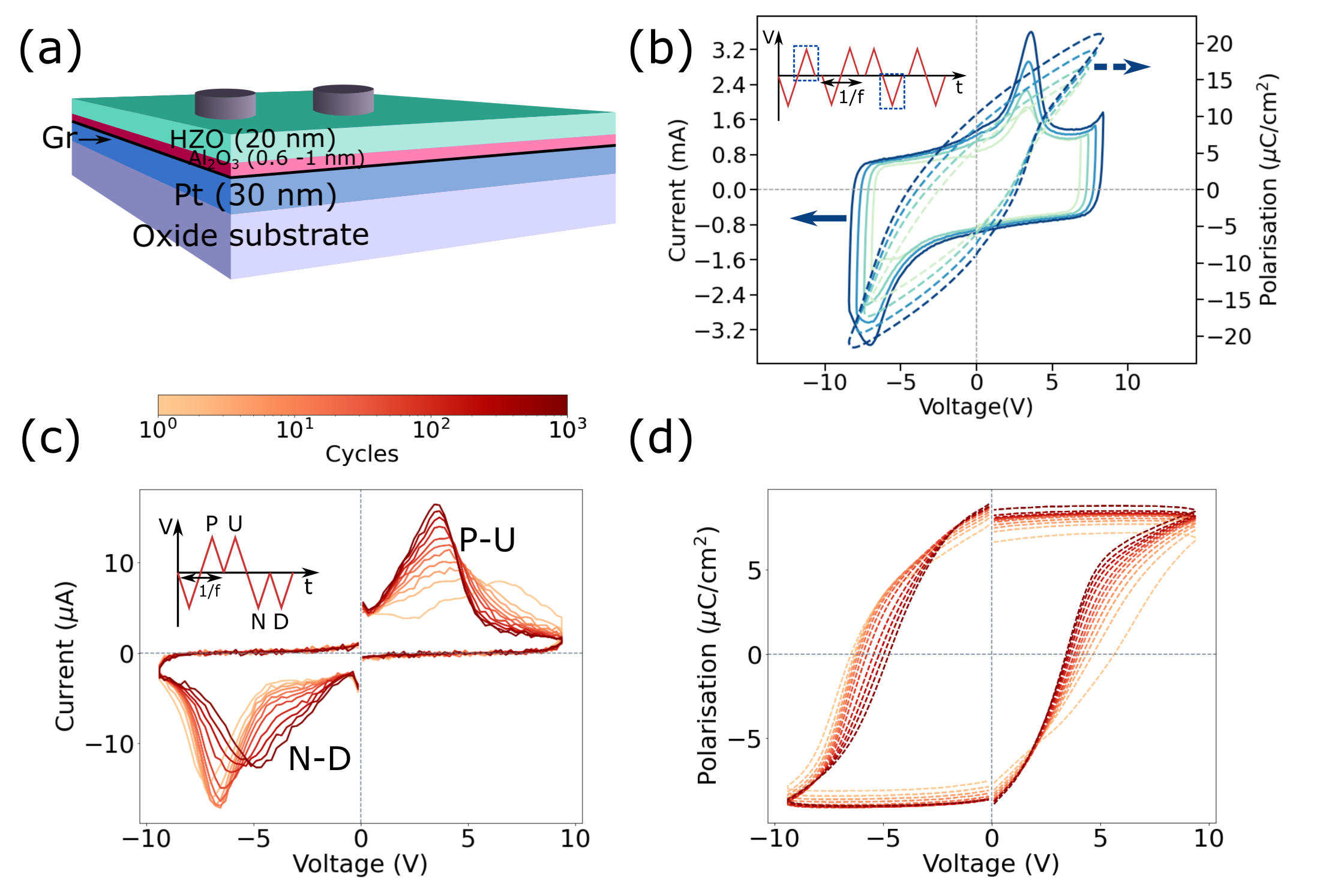}
	\captionsetup{font=footnotesize}
	\caption{Electrical characterization of HZO films: (a) schematic and  (b) measured switching characteristics of devices with 20\,nm HZO deposited after 6 cycles \AlOx{} for nucleation (inset: DHM pulse sequence, with blue dashed lines highlighting the pulses plotted here); (c) current-voltage curves (from PUND measurements, pulse train shown inset) and (d) calculated polarization-voltage curves with cycling. Devices were cycled at 10\,V, 100\,kHz and measured at 10\,V, 10\,kHz.}
\label{fig2_AlOx}
\end{center} 
\end{figure}

After observation of the o-phase in the deposited films, electrical polarization-voltage (P-V) and current-voltage (I-V) measurements were performed for confirmation of their ferroelectricity, as presented in figure \ref{fig2_AlOx}. The top Al$_2$O$_3$ layer used for crystallization was etched away for HZO films of 20\,nm thickness, and capacitors were defined with diameters of 200 $\upmu$m (figure \ref{fig2_AlOx}(a); full details are given in the supplementary materials). The I-V (solid lines) and P-V (dashed lines) characteristics with increasing applied voltage are shown in figure \ref{fig2_AlOx}(b). Dynamic hysteresis measurements (DHM) were performed using the pulse train shown in the inset. The voltage was varied from 7-8.5\,V. There is a pronounced imprint (shift to negative values along the voltage axis), and at lower voltages the negative switching peak can't be measured.

At the maximum voltage applied for DHM (8.5\,V), the coercive voltages for the stack are \textit{V$_c^{+/-}$} = +2.7/-4.5 V and the maximum measured remanent polarization is \textit{P$_r^{+/-}$} = +10.3/-8.9 \PrUnits, with a lower Pr in the negative polarity due to the cut-off of the negative switching peak. Following literature, a 2Pr of 19.2 \PrUnits{} in the pristine state  corresponds to an o-phase fraction of around 40\%,\cite{kashir2021defect} which fits well to the phase fractions obtained by GIXRD fitting. The high coercive voltages are related to the addition of the \AlOx{} layer to the stack \cite{lomenzo2019ferroelectric} and indicate a very low dielectric constant for the Al$_2$O$_3$, which is likely due to incomplete ALD reactions leading to the incorporation of precursor residuals in the \AlOx{} film \cite{zheng2014improvement}. 

Electrical measurements performed with the crystallization layer intact are shown in the supplementary. Reducing the ferroelectric thickness from 20\,nm to 10\,nm increases \textit{2Pr} from 6.4 \PrUnits{} to 15.66 \PrUnits{}. However, this is to the detriment of the film endurance, and full switching couldn't be measured on 10\,nm devices. Etching of the top layer further reduced the film endurance so that no switching could be measured.

Figures \ref{fig2_AlOx}(c) and \ref{fig2_AlOx}(d) show PUND (Positive Up, Negative Down, see pulse train inset) cycling measurements on 20\,nm films nominally identical to those discussed above. This method allows us to separate switching from non-switching contributions to the current, i.e. the dielectric displacement current and leakage. The layers show a \textit{2Pr} around $\sim$ 20\,$\mu$C/cm$^2$, consistent with the previous measurements, and switching persists for 10$^3$ cycles before device breakdown.

To summarize the results on in-situ Al$_2$O$_3$ nucleation, the endurance was found to be highly variable, with some devices showing large defects, and large coercive fields were observed for all samples. This points to a poor structure and low dielectric constant of the deposited Al$_2$O$_3$ layer \cite{zheng2014improvement}. In order to have reliable integration of HZO on the in-situ ALD layer, one possibility is increasing the conformality of the \AlOx{} layer by careful control of the precursor pulse and purge times, and deposition temperature \cite{zheng2014improvement, zhang2014direct}. Nonetheless, the thickness of the nucleation layer is clearly critical in increasing the device coercive voltage, and in general, deposition of conformal \AlOx{} on Gr requires relatively thick oxides of several nms \cite{vervuurt2017uniform}. This will also lead to a destabilization of the ferroelectric state due to depolarization fields \cite{lomenzo2019ferroelectric}. 

\section{Ex-situ methods for nucleation on graphene}
\subsection{Film deposition and characterization}

\begin{figure}[htb]
\begin{center}
		\includegraphics[height=6cm]{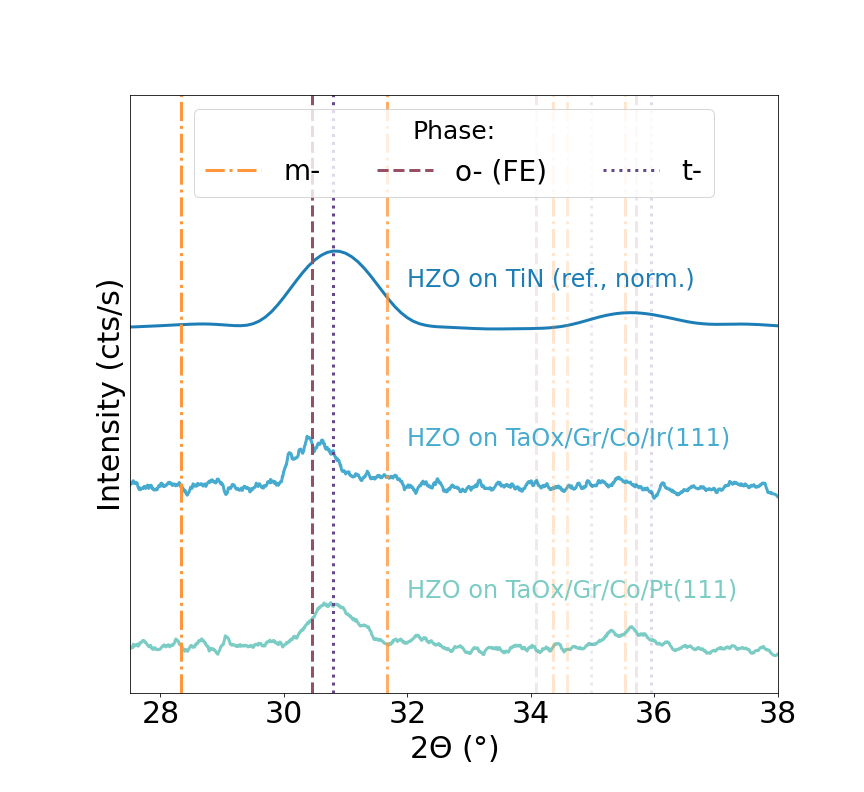}
	\captionsetup{font=footnotesize}
	\caption{GIXRD characterization of ex-situ (TaOx) nucleated HZO films: 2$\Theta$ scans of HZO films deposited on two different stacks with Ta as a nucleation layer}
\label{fig3_Ta}
\end{center} 
\end{figure}

To that end, another approach was investigated, where the nucleation layer for HZO is an ex-situ deposited transition metal which oxidizes in atmosphere before the ALD process. After the Gr/HM stack processing, an additional 2-3\,nm layer of Ta was deposited to form the nucleation layer. Previous transmission electron microscopy measurements, not shown, indicate that the thin top layer of Ta becomes oxidized upon exposure to atmosphere. Then, 11\,nm HZO and 2\,nm \AlOx{} were deposited and annealed as before. GIXRD measurements on these samples are presented in figure \ref{fig3_Ta}.

Ferroelectric o-/t-phase peaks were observed in GIXRD scans, as shown in figure \ref{fig3_Ta}. Compared to the film deposited on \AlOx{}, the non-ferroelectric m-phase is reduced, but still larger than in reference samples with standard TiN electrodes. As oxygen vacancies are critical for stabilizing the ferroelectric o-phase \cite{hoffmann2015stabilizing}, more oxygen vacancies are likely generated in the film with TaOx as a nucleation layer, and strong oxygen scavenging at the bottom interface in contact with TaOx should lead to the formation of a tetragonal interfacial layer \cite{cheng2022reversible}. On the other hand, the top \AlOx{} capping layer will form a barrier for oxygen scavenging at the top electrode during annealing, somewhat reducing the o-phase fraction compared to the reference HZO with TiN electrodes. Based on GIXRD fittings (see supplementary information), we estimate that the full 2Pr one could expect to observe in these films is around 26 \PrUnits{} \cite{mukundan2020quantifying} in the case of Pt(111), and somewhat lower for Ir(111).

Also visible in figure \ref{fig3_Ta} is a significant shift of the HZO peaks, which indicates an altered strain state in films deposited on stacks with a different heavy metal element (blue: Ir/ green: Pt). Strain may modify the grain size and thereby the phase of HZO \cite{materlik2015origin}. These observations demonstrate that the underlying support influences the crystal properties of deposited HZO, which offers a path for tuning the ferroelectric properties. As discussed in the supplementary materials (see supplementary information), GIXRD peak positions are similar for the reference sample with TiN electrodes and the stack with Pt(111); furthermore, this stack appears to show a higher phase fraction of ferroelectric o-phase (see supplementary information). Therefore the sample on Pt(111) was chosen for electrical investigation of the FE properties.

\begin{figure}[htb]
\begin{center}
		\includegraphics[width=15cm]{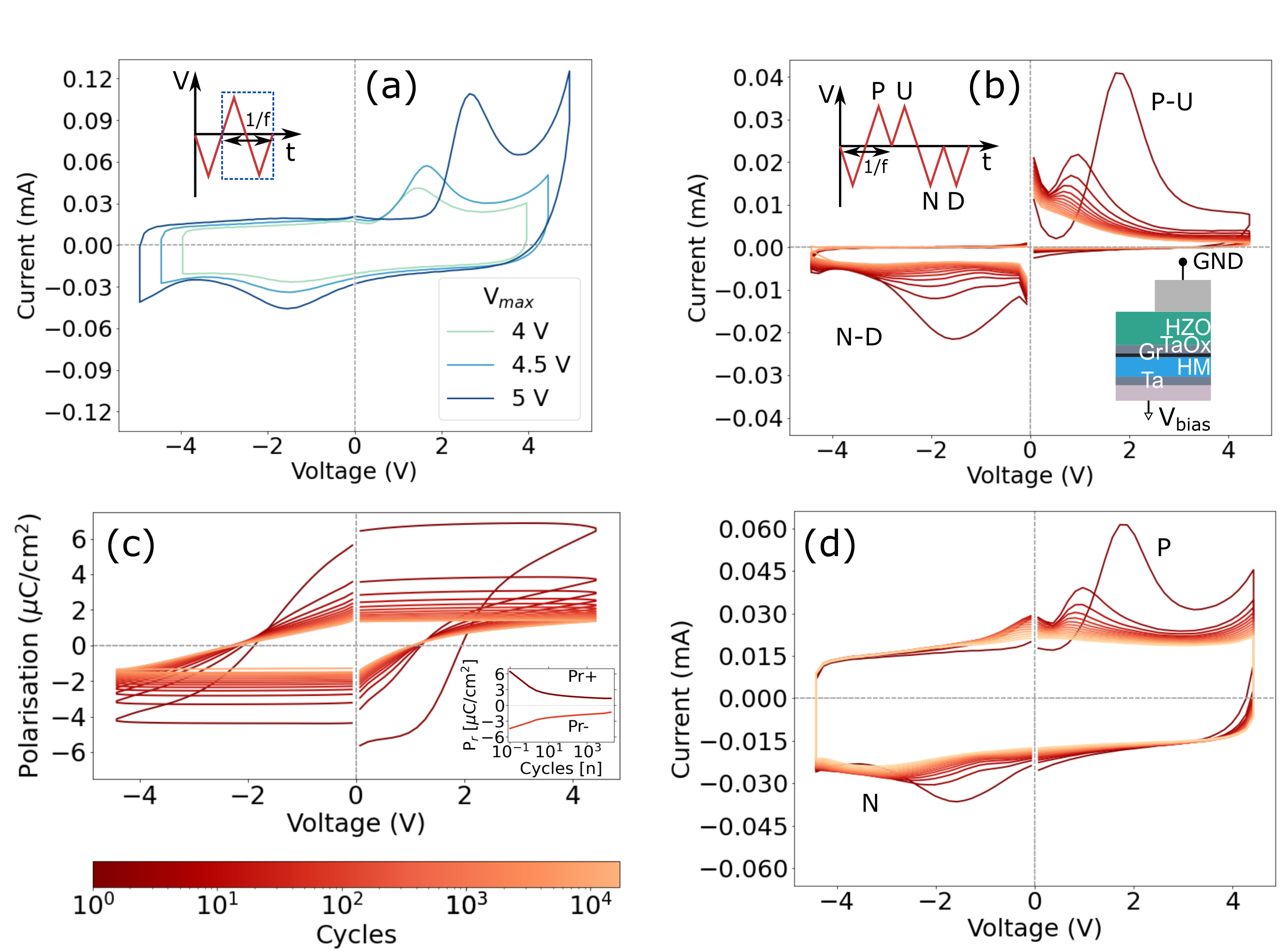}
	\captionsetup{font=footnotesize}
	\caption{Electrical characterization of films deposited on TaOx interlayer:  (a) IV switching curves on uncycled, prepoled capacitors (inset: pulse train, with blue dashed lines indicating plotted pulses), for the Pt(111) sample; (b) I-V measured on the same sample with the PUND method over 10$^4$ switching cycles (inset, top: PUND pulse train; bottom: stack schematic, indicating direction of applied bias); (c) P-V endurance curves found by integrating the switching currents in (b) (inset: change in Pr as a function of cycles); (d) I-V measured on the P and N pulses only.}
\label{fig4_TaElec}
\end{center} 
\end{figure}

\subsection{Electrical characterization of ex-situ nucleated films}

For electrical measurements, the \AlOx{} capping layer used for crystallization was etched away before defining capacitor structures via deposition of Ti/Pt electrodes (see supplementary for details). Electrical characterization of these capacitors are presented in figure \ref{fig4_TaElec}. First, switching peaks with increasing maximum voltage $V_{max}$ were applied consecutively to the same sample, as shown in the switching IV curves shown in \ref{fig4_TaElec}(a). Switching peaks were observed at a maximum applied voltage \textit{V$_{max}$} of \textless 5 V, with a coercive voltage which is already greatly improved compared to the Al$_2$O$_3$ nucleation layer. This is due to the larger dielectric constant of Ta$_2$O$_5$ and subsequent smaller voltage drop over the interlayer. In the positive direction, a significant leakage current is observed at \textit{V$_{max}$} = 5 V, which limits the possible maximum applied voltage for electrical cycling. At the same time, the switching distributions are strongly asymmetric in both polarities. This asymmetry can originate from the TaOx layer modifying the local field distribution and domain nucleation probability depending on the polarity of the applied voltage.

Switching measurements with intermediate cycling, otherwise known as endurance measurements, were performed at 4.5\,V and a measurement frequency of 1\,kHz. Cycling was performed with square pulses at 100\,kHz. For these measurements, a PUND pulse train, as shown inset in figure \ref{fig4_TaElec}(b), was applied between cycles. The bias was applied to the bottom electrode in contact with the TaOx layer (also shown inset in figure \ref{fig4_TaElec}(b)). 

Figure \ref{fig4_TaElec}(c) shows the extracted P-V curves with cycling, where capacitors display ferroelectric behavior with \textit{P$_r^{+/-}$} up to +6.44/-4.37 \PrUnits{} and \textit{V$_c^{+/-}$} of +2.02/-1.87 V. However, the polarization decreases with cycling, known as fatigue. Fatigue is common in HZO films, but is pronounced in these devices, with the switching completely lost after about 50 switching cycles (see the change in Pr with cycling, shown inset in figure \ref{fig4_TaElec}(c)). Furthermore, no wakeup is observed, i.e. the polarization switching is not seen to improve at the onset of cycling. The fatigue corresponds with a shift in \textit{V$_c$} to more negative voltages, i.e. a negative imprint. When plotting the uncorrected ´P' and ´N' pulses (figure \ref{fig4_TaElec}(d)), a switching peak at 0\,V becomes visible, particularly pronounced in the positive direction. The combination of imprint and switching at 0\,V signals a strong depolarization field and the build-up of an internal bias which favors one ferroelectric state over the other.

When switching into the negative direction, the applied electric field is not large enough to fully switch back all domains, due to asymmetric field distributions and/or possible structural changes. The negative imprint points to a build-up of positive charges at the bottom interface (where the bias is applied), which will facilitate switching into the positive polarity over switching into the negative polarity and leads to a pinning of domains with polarization pointing towards the top electrode. Positive charges in such ferroelectric bilayer stacks can generally be attributed to oxygen vacancies, which can be generated by oxygen scavenging by the substoichiometric TaOx layer. The build-up of oxygen vacancies at the bottom electrode could additionally lead to the formation of a tetragonal interfacial layer, which generates a depolarization field \cite{cheng2022reversible} and further reduces the effective field over the ferroelectric. This process is repeated cycle-by-cycle, thus quenching the Pr with cycling. There is an absence of such quenching effects when increasing the field during subsequent measurements on the same capacitor (figure \ref{fig4_TaElec}(a)), implying that the fatigue is reversible under higher fields \cite{cheng2022reversible, gong2021observation}.

\begin{figure}[htb]
\begin{center}
		\includegraphics[width=15cm]{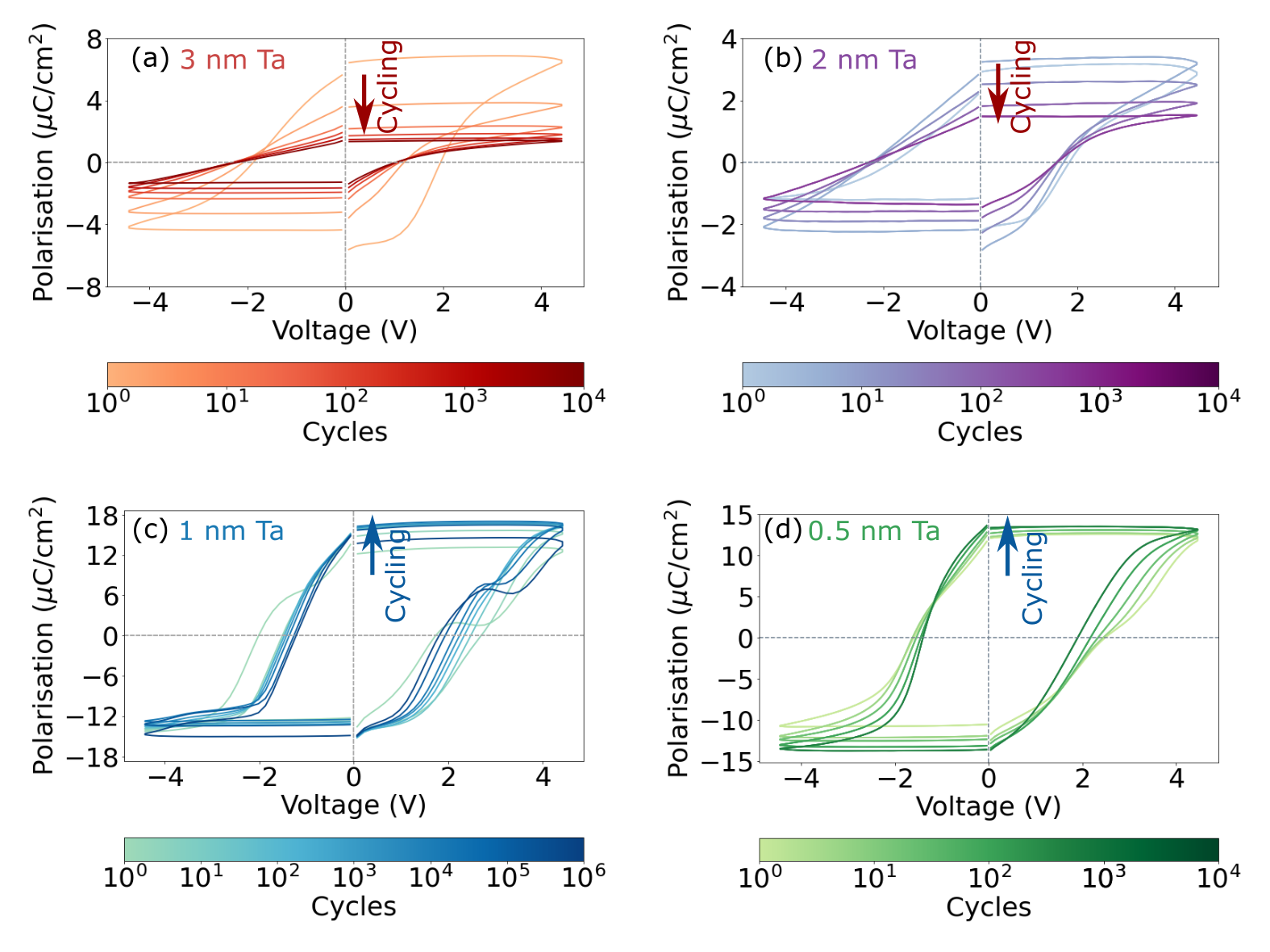}
	\captionsetup{font=footnotesize}
	\caption{Electrical characterization of stacks with varying TaOx interlayer thickness:  Polarization-voltage endurance curves of \HZO{} deposited stacks with Ta interlayers (oxidized in air) of (a) 3\,nm; (b) 2\,nm; (c) 1\,nm; (d) 0.5\,nm. 
	Devices are cycled at 100\,kHz and measured at 1\,kHz. The arrows indicate fatigue or wakeup during the initial polarization cycles.}
\label{fig5_TaOx}
\end{center} 
\end{figure}

To confirm this, a sample series was made where the thickness of the top TaOx layer was systematically reduced. The samples used for these measurements consisted of Gr/Pt (10\,nm), with deposited top Ta layers of varying thickness, partially oxidized by exposure to ambient conditions. In figure \ref{fig5_TaOx}, P-V curves extracted from PUND measurements are presented for varying TaOx interlayer thicknesses. It can be clearly seen that for thicker nucleation layers (figures \ref{fig5_TaOx}(a-b)), the polarization decreases immediately upon field cycling. On the other hand, when the TaOx thickness is reduced to 1\,nm and below (figures \ref{fig5_TaOx}(c-d)), the films display a typical ferroelectric wake-up behavior, with 2Pr values up to 28\,\PrUnits{}, consistent with estimates from GIXRD. Looking at the cycling behavior, we see that the fatigue is slower for the 2\,nm interlayer sample than 3\,nm, while the maximum Pr and endurance peak in the 1\,nm interlayer sample before decreasing slightly for 0.5\,nm. 

This can be explained by the thinner interlayers becoming fully oxidized during the sample transfer and in the first ALD cycles, whereas thicker interlayers act as an oxygen sink or scavenging layer during electric field cycling. Similar to the case of supplying higher or lower oxygen doses during ALD \cite{schroeder2019recent}, the oxidation state of the scavenging layer influences the endurance and Pr of the films, while the interlayer thickness influences the magnitude and rate of polarization loss \cite{lancaster2022investigating}. This should be further controllable by varying the oxygen supplied during the ALD process, to reach optimal ferroelectric properties.  

\section{Post-deposition sample characterization}
\subsection{Physical characterization of HZO on Gr/HM stacks}
Finally, the impact of the ALD process on underlying Co/Pt(111) layers was investigated. During the stack deposition process, the Co layer is intercalated underneath the Gr layer by heating at a defined temperature. However, high temperatures can lead to intermixing of the Co/Pt(111) and a subsequent reduction in magnetic hysteresis \cite{vermeulen2019ferroelectric} and DMI \cite{zimmermann2018dzyaloshinskii}. The Gr/Co/Pt(111) fabrication process is described fully in \cite{ajejas2018unraveling, blanco2021large} and the supplementary materials, and the temperature dependence of intermixing was investigated in \cite{ajejas2019thermally}. Here, the effect of the ALD process on the structural and magnetic properties of the Co/Pt(111) bilayers is investigated. 

\begin{figure}[!htb]
    \centering
        \includegraphics[height=10cm]{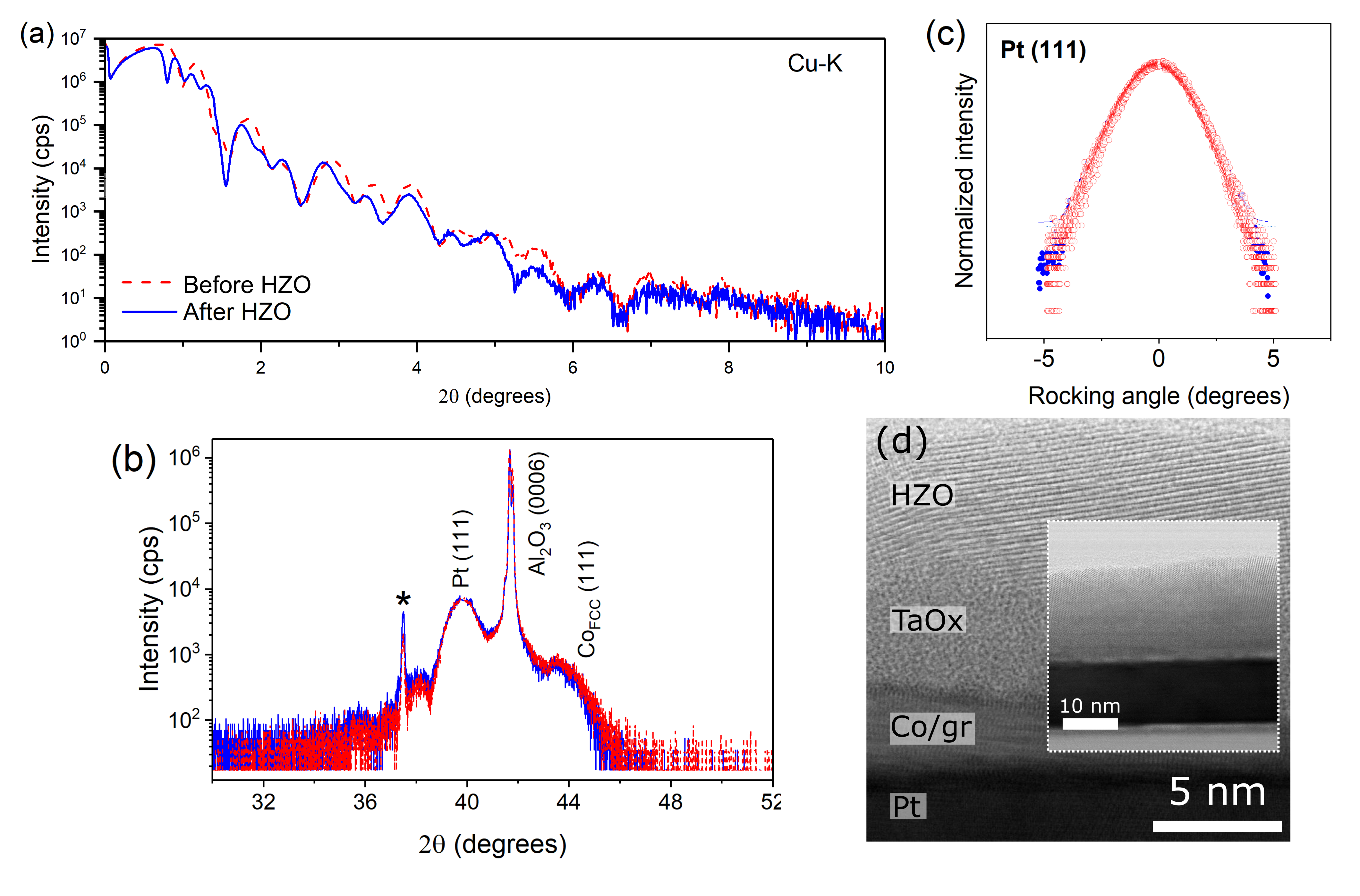}
   \caption{Physical characterization of the films before and after HZO deposition: (a) XRR measurements and (b) $\theta-2\theta$ diffraction patterns recorded in the Gr-based heterostructures before and after the HZO deposition. (c) Rocking scans around the Pt(111) reflection before and after ALD. Sharp contribution at 37.5º(*) corresponds with Al$_2$O$_3$[0006] reflection at Cu K-$\beta$ energy. (d) high resolution annular bright field STEM image of the stack with 3\,nm Ta interlayer. Inset: larger-area STEM image showing conformal HZO deposition.}
    \label{x_ray}
\end{figure}

X-Ray Reflectivity (XRR), X-ray diffraction (XRD) $\theta-2\theta$ measurements and rocking-scans were performed in a parallel beam geometry on a commercial Rigaku SmartLab SE multipurpose diffractometer. Figure \ref{x_ray} shows the X-ray analysis of the film stacks before and after HZO deposition. In figure \ref{x_ray}(a), XRR measurements show the appearance of well-defined intensity oscillations due to the interference between reflected x-rays at the different interfaces of the heterostructure, demonstrating the presence of abrupt interfaces, also after HZO deposition.
Besides the Al$_2$O$_3$[0006] crystallographic reflection from the substrate (at $2\theta=41.7\degree$), $\theta-2\theta{}$ XRD patterns in figure \ref{x_ray}(b) recorded in the heterostructures present an intense peak centered at $2\theta=39.8\degree$, which corresponds with Pt[111]. Despite the reduced Co thickness, we could observe a further peak at $2\theta=44.2\degree$ due to the relaxed out-of-plane Co$_{fcc}$[111]. Figure \ref{x_ray}(c) presents the rocking curves around the Pt[111] reflection proving that HZO deposition does not induce additional structural disorder in the Pt structure.

The XRD measurements demonstrate the epitaxial growth of the Pt and Co layers, which grow (111)-oriented on top of the Al$_2$O$_3$ (0001) substrate. After HZO deposition, contributions from Pt and Co remain unchanged, indicating the preservation of the structural properties of the stack after the ALD proccess.

A high-resolution annular bright field STEM image of the stack deposited onto Pt(111) is represented in figure \ref{x_ray}(d). The HZO layer is polycrystalline, while the TaOx capping layer is amorphous. The measurements were performed in a JEOL ARM200cF microscope after preparing cross-sections of the samples by mechanical polishing and Ar ion milling. Low magnification images (inset, also shown in higher resolution in the supplementary information) exhibit continuous layers over long lateral distances. No major interfacial intermixing due to the processing of the FE layer can be detected from these images. The chemical composition of the films was measured with electron energy-loss spectroscopy (EELS, see supplementary information), further confirming that the deposition method leaves the underlying stack intact.

\subsection{Magnetic hysteresis measurements on HZO on Gr/HM stacks}

The magnetic properties of the films with a 3 nm TaOx interlayer were ex-situ investigated at RT by means of magneto-optical Kerr effect (MOKE) magnetometry in polar geometry. In precedent works, we have demonstrated that on both Pt and Ir HM buffers, the samples present a well-defined PMA for Co layer thinner than 4 and 2 nm for Pt and Ir buffers, respectively \cite{blanco2021large}. For low Co thicknesses, the hystereses present sharp transitions and a remanence magnetization (i.e., at zero field) almost 100\% of the magnetization saturation value ($M_S$, i.e. at field larger than the anisotropy field), as well as large coercive fields. For larger thickness, the reduction of both magnetization remanence and coercive field together with the appearance of smoother magnetization  reversal transitions, indicate a spin reorientation from out-of-plane to in-plane magnetic anisotropy. By combining Kerr, X-ray magnetic circular dichroism, high-resolution transmission electron microscopy and ab-initio modelling, we previously ascribed such behavior to a large magnetic anisotropy due to the anisotropy of the orbital moment, which is of interfacial nature \cite{blanco2021large}.

\begin{figure}
    \centering
        \includegraphics[height=6cm]{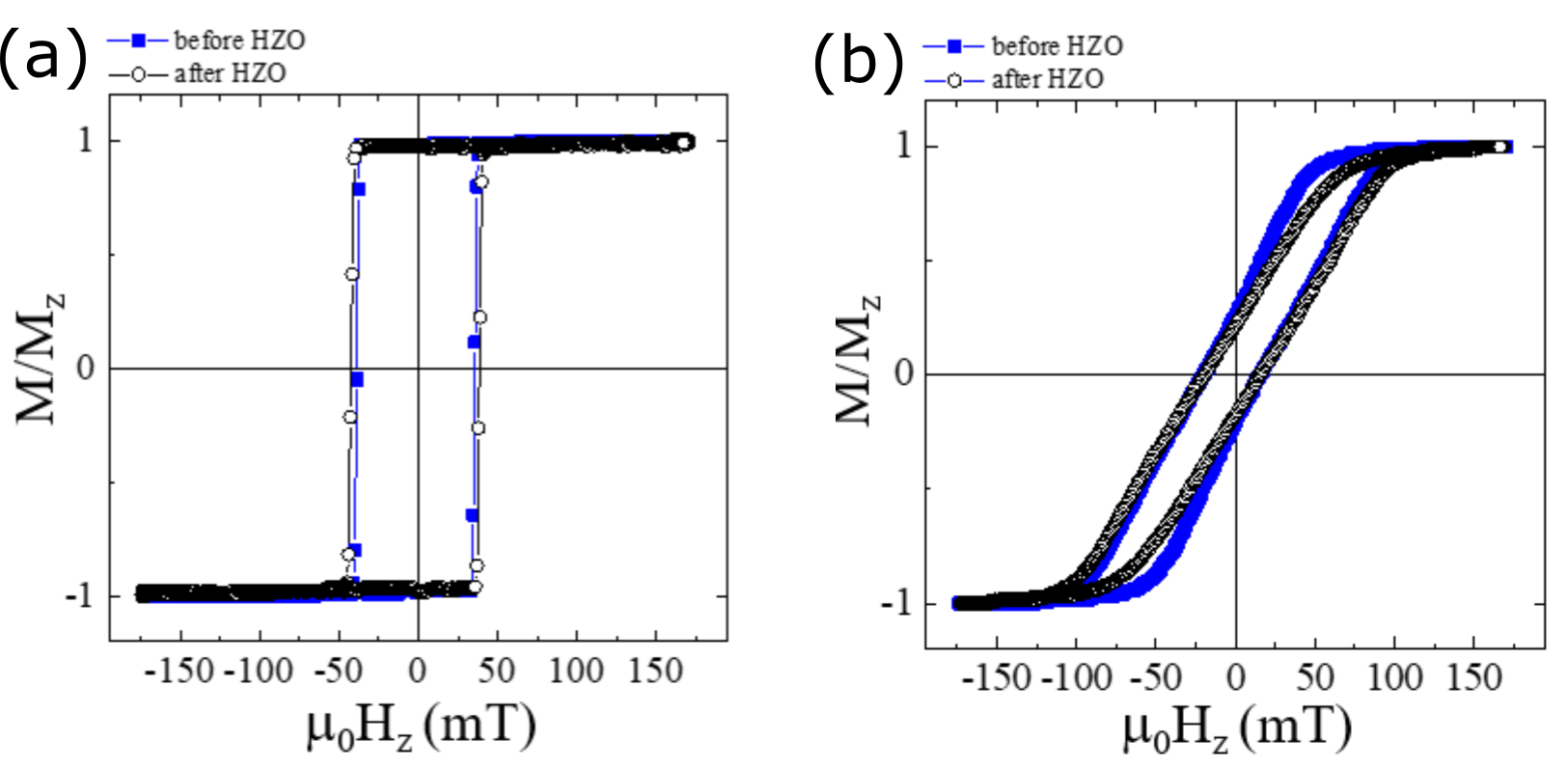}
   \caption{Magnetic hysteresis measurements on full stacks: Out-of-plane magnetization component $M_z$ normalized to the saturation value $M_S$ as a function of the applied perpendicular magnetic field $\upmu_0$H$_z$ acquired by magneto-optical (polar) Kerr magnetometry before (blue filled squares) and after (empty black circles) the ALD deposition of the ferroelectric HZO for: (a) a 2 nm Co layer between Gr and Pt(111) and (b) a 4 nm Co layer between Gr and Pt(111). In all cases, the HZO was deposited on a TaOx interlayer.}
    \label{mag_hyst}
\end{figure}

Figure \ref{mag_hyst} shows the out-of-plane magnetization component $M_z$ normalized to the saturation value $M_S$ as a function of the applied perpendicular magnetic field $\upmu_0$H$_z$ acquired before and after the ALD deposition of the ferroelectric HZO layer. Measurements demonstrate that the samples possess PMA both without (blue filled squares) and with (empty black circles) the HZO layer for two selected Co thickness, i.e. 2 and 4 nm, on Pt buffers.
The cycles measured before (blue filled squares) and after (empty black circles)  the HZO deposition in the stacks with Pt-buffers in figure 7(a) and (b) are similar and do not present relevant difference, while PMA is maintained. As discussed in reference \cite{blanco2021large}, PMA is not achievable in Co/Pt bilayers at the Co thicknesses here discussed, except in the presence of Gr. Thus, these results indicate that the Gr stays intact during the ALD deposition and that the magnetic characteristics of the Gr/Co/HM stacks behave as expected. This is in good agreement with recent studies showing that an interlayer can successfully protect Gr from damage during ALD with O$_2$ plasma \cite{canto2021plasma}. 

Nonetheless, in the case of Gr/Co/Ir(111) (see supplementary information), we notice a slight increase of coercive field after the ALD deposition of the HZO, which is probably due either to a minor partial oxidation of Co or Co/HM intermixing (this may be favored by the elevated deposition temperature used for the ALD growth and subsequent annealing, to be further investigated). This effect is greatly reduced in Gr/Co/Pt(111) samples (figure \ref{mag_hyst}(a) and \ref{mag_hyst}(b)). In any case, the strong similarity in PMA before and after deposition indicates that these are both minor effects, which do not substantially affect the overall magnetization reversal behavior of the Gr/Co system. 

\section{Conclusions}
Deposition of ferroelectric HZO on graphene was demonstrated utilizing both an in-situ and an ex-situ nucleation layer method. The optimized in-situ layer was 6 cycles of \AlOx{} deposited at 150$\degree$C with H$_2$O as an oxidant, providing good wetting of the graphene layer which enabled subsequent film deposition with \textit{2P$_r$} values up to $\sim$ 20 \PrUnits. However, relatively thick \AlOx{} layers are needed to ensure conformal deposition, leading to high coercive voltages. An ex-situ deposited Ta film, oxidised in atmosphere, provides uniform coverage and an adequately reactive surface for the deposition of HZO. In this case, the films could be switched at \textless 5 V, with \textit{2P$_r$} values varying up to 28\,$\mu$C/cm$^2$ for 1\,nm TaOx interlayers. 

Varying the interlayer thickness could modify the ferroelectric properties and change the device behavior from immediate fatigue, to wakeup and stable polarization switching. From this we hypothesize that the mechanism for fatigue in samples with thicker interlayers is oxygen transfer from the HZO to the TaOx interlayer during cycling. Further control over the oxygen gradient and a reduction in the scavenging effect could come from a stoichiometric Ta$_2$O$_5$ layer, combined with modifying the oxygen content of the HZO film itself \cite{schroeder2019recent}. 

Through utilizing a Gr/ferromagnet/heavy metal support we find that the strong interfacial perpendicular
magnetic anisotropy remains almost unaltered upon the ALD deposition of the HZO layer, as demonstrated by the magnetization vs. field hysteresis measurements. This proves that the ALD process leaves the Gr layer intact, as the ferromagnetic Co layer would be completely oxidized without the protection of Gr and hence the magnetization hysteresis loops would not be observed.

These first results open up pathways for investigating the integration of HZO and other \HfO-based ferroelectrics directly on graphene. Combined with recent advances in the CVD growth of graphene on insulating substrates \cite{sun2016direct} or on semiconductor materials such as Ge,\cite{lukosius2016metal} this could enable the transfer-free processing of graphene as a channel material in non-volatile memories and various spintronic devices. The high remanent polarization demonstrated at room temperature in HZO/graphene systems, combined with the observed magnetic hysteresis, holds promise for spin-orbit logic and skyrmionic devices. Finally, the deposition method is versatile and can be applied to graphene fabricated using different processes on various substrates.

\section{Experimental section}
\subsection{Chemical Vapor Deposition of Graphene on Pt(111)} To obtain high-quality epitaxial Gr we utilized a chemical vapor deposition (CVD) method in ultra-high-vacuum (UHV) condition \cite{ajejas2019thermally}. In constrast to exfoliation methods, we followed a UHV growth and in-situ characterization procedure, enabling the electronic and chemical control of the Gr interfaces while avoiding contamination, which are known to affect the overall transport properties. 
The fabrication of the samples was monitored by in-situ surface analysis at each stage of the growth, which consisted first in the deposition of a 2 nm Ta sticking layer and 10 nm-thick, epitaxial, single-crystal, heavy-metal buffer (i.e., Pt(111) or Ir(111)) by dc sputtering at 550 K onto commercial SrTiO3(111) single crystal substrates. To note that a 2nm layer of Ta on top of oxide substrate was used as a sticking layer before the deposition of Pt(111) when the latter has been set below 10nm. 
The HM buffers present a crystal quality equivalent to Pt(111) and Ir(111) single crystals, as verified by x-ray diffraction measurements (not shown). \\

Then, epitaxial monolayer graphene was grown in-situ by ethylene chemical vapour deposition at 1025 K on top of the buffers. In the case of the Pt(111) buffer, the LEED and STM analyses indicate the presence of multi-domain Gr flakes (i.e., domains oriented ±15º). Instead, Gr grown on the Ir(111) buffers presents the well-known “10x10” moiré pattern that is due to the incommensurate unit cells of Ir and Gr. The difference between the two cases is mostly related to the different interaction of Gr with the two heavy metals, Gr-Ir being  the more intense. The details of the growth and characterization of the epitaxial stacks are reported in \cite{ajejas2019thermally}. 
The epitaxial Gr/HM(111) onto SrTiO3(111) systems are then used for the ferromagnetic Co layer intercalation, as discussed in the following section.

\subsection{Ta(Ox)/Graphene/Co/HM stack deposition} To incorporate a ferromagnetic layer into the epitaxial system described above, we resorted to the thermally activated intercalation method as reported in \cite{ajejas2019thermally}. In brief, Co was evaporated on top of Gr by molecular beam epitaxy (MBE) at RT with a deposition rate of 0.3 \AA/s monitored by an in situ quartz balance. Then, in order to favor the penetration of Co atoms underneath the Gr layer, a low-temperature anneal was applied to activate the intercalation process. Surface microscopy and in-situ spectroscopy experiments revealed that the intercalated Co is pseudomorphic with the metallic substrates, epitaxial and (111)-oriented, as well as homogeneous (over $\sim\upmu$m2) underneath Gr. In addition, these studies demonstrated that i) the Gr efficiently protected the Co from oxidation, and ii) at annealing temperature higher than 600 and 650 K for Pt and It respectively Co/HM intermixing occurs. \\

Finally, for some samples, we deposited by in-situ dc sputtering at RT 2 nm thick Ta layer on top of the final structure in order to investigate the HZO layer nucleation on Gr and to protect the underlying Co from oxidation.

\subsection{Atomic Layer Deposition of \HZO/\AlOx} Films were deposited in an Oxford OpAL ALD system. First, where applicable, \AlOx{} as a nucleation layer was deposited at 150\degree C with TMA as the precursor and H$_2$O as the oxidant. Next, the temperature was ramped up to 250\degree C, for deposition of the HZO and subsequent \AlOx{}, with HyALD/ZyALD/TMA as the Hf, Zr and Al precursors, respectively. For HZO films, the oxidant used was a remote O$_2$ plasma, at a power of 300 W and with the source located $\sim$30 cm from the substrate; for \AlOx, H$_2$O was employed as the oxidant. The HZO films were deposited at a layer ratio of 1:1 with an approximate growth rate of 1 \AA/cycle. HyALD and ZyALD pulse times were 2s, increased to 3s during the first nucleation experiment. The HZO thickness was varied by changing the number of cycles, while the \AlOx{} capping layer for crystallization was kept constant at a nominal thickness of 2 nm. For the TaOx thickness series, instead of \AlOx{}, a 10\,nm TiN top electrode was sputtered under high vacuum. \\

Samples were annealed for 5 s at 600\degree C in an AST SHS-2800 RTP under N$_2$ atmosphere for crystallization of the HZO.

\subsection{Atomic Force Microscopy measurements} AFM was performed in atmosphere on a Bruker Dimension XR microscope in tapping mode, with a Nanosensors Point-Probe PPP-NCHR tip of 7 nm radius. 

\subsection{X-Ray Diffraction measurements} For GIXRD, 2$\theta$ scans were measured in a Bruker D8 Discover XRD system with a Cu K-$\alpha 1$ X-ray source, a G{\"o}bel mirror and 0.2 mm divergence slit on the primary side, and a 2.5\degree{} Soller slit on the secondary side. The source was set to an angle $\Omega$ = 0.45\degree{} to the sample surface. \\
X-Ray Reflectivity (XRR), X-ray diffraction (XRD) $\theta-2\theta$ measurements and rocking-scans were performed in a parallel beam geometry by using a commercial Rigaku SmartLab SE multipurpose diffractometer equipped with a Cu K-$\alpha$ source ($\gamma = 0.154$\,nm), a cross beam optics system and a D/Tex Ultra 250 1D silicon strip detector. The measuremnts were performed with a 2.5\degree{} Soller slit and a 2 mm length limit slit at the primary side and a second 2.5\degree{} Soller slit at the secondary side.

\subsection{Capacitor fabrication} 10\,nm Ti and 25\,nm Pt were deposited via e-beam evaporation under high vacuum through a shadow mask to form dots for the capacitors. Capacitors were either deposited on the \AlOx{} layer, or directly to the HZO by first etching the samples for 90 s in AZ 300 MIF developer at room temperature before metallization (for \AlOx{} capping layers) or in SC-1 etch H$_2$O:H$_2$O$_2$:NH$_4$OH 5:1:1 at room temperature (for TiN capping layers). 

\subsection{Electrical characterization} DHM measurements were performed on an Aixacct TF3000 ferroelectric prober, at a frequency of 10 kHz. \\

IV switching and PUND measurements were performed using PMUs controlled by Keithley 4225 RPMs on a Keithley 4200 semiconductor probe station. The frequency for both switching and PUND measurements was 1\,kHz, and no time delay was applied between PUND pulses. In endurance measurements, the device was cycled with square pulses at 100\,kHz at the measurement voltage amplitude. \\

\subsection{Scanning transmission electron microscopy and electron energy-loss spectroscopy} Electron microscopy observations were carried out in a JEOL ARM200cF microscope equipped with a CEOS spherical aberration corrector and a Gatan Quantum EEL spectrometer at the Centro Nacional de Microscopıa Electronica (CNME) at the University Complutense of Madrid. Specimens were prepared by conventional methods, including mechanical polishing and Ar ion milling. Energy-loss electron spectroscopy (EELs) maps are created by a multiple linear least squares fitting in order to integrate at each edge of interest and to differentiate between overlapping edges. Data are plotted directly from the integrated intensity, which is proportional to the thickness of the material. It should be noted that due to the sample preparation, this thickness is inhomogeneous and as such the chemical composition is not quantitatively mapped. \\

\subsection{Polar Kerr Magnetic hysteresis measurements} The RT vectorial-Kerr experiments were performed in polar configuration by using p-polarized light (with 632\,nm wavelength) focused on the sample surface and analyzing the two orthogonal components of the reflected light as function of the magnetic field applied along the sample out-of-plane ($\hat z$) direction. This enables the acquisition of the hysteresis loops of the out-of-plane magnetization components, $M_Z$, which has been normalized to the magnetization saturation $M_S$.   

\medskip
\textbf{Supporting Information} \par 
Supporting Information is available online. 

\medskip
\textbf{Acknowledgements} \par 
The authors would like to acknowledge fruitful discussions with Vincent Cros, Pierre Seneor, Bruno Dlubak and Nicolas Reyren from CNRS-Thales.

This project has received funding from the FLAG-ERA JTC 2019 grant SOgraphMEM through the partner’s national research agencies AEI/MICINN (Spain, PCI2019-111867-2) and DFG (Germany, MI 1247/18-1). 
IMDEA team acknowledges support by the Community of Madrid (CM) through Project P2018/NMT-4321 (NANOMAGCOST), by MICINN through Projects RTI2018-097895-B-C42, 43 (FUN-SOC) and PID2021-122980OB-C51,52 (ECLIPSE), and by the ’Severo Ochoa’ Programme for Centres of Excellence in R\&D CEX2020-001039-S. AG and IA ackowledge support from CM (PEJD-2017-PREIND-4690 and PEJD-2019-POST/IND-15343) and JMD from MICINN (BES 2017-080617). 

{\footnotesize
\bibliography{Manuscript}
\bibliographystyle{unsrt}}

\end{document}


\maketitle 
\newpage
\section{AFM analysis of Gr surface after ALD deposition without an interlayer}
\begin{figure}[h]
\begin{center}
	\subfloat[]{
		\includegraphics[height=5cm]{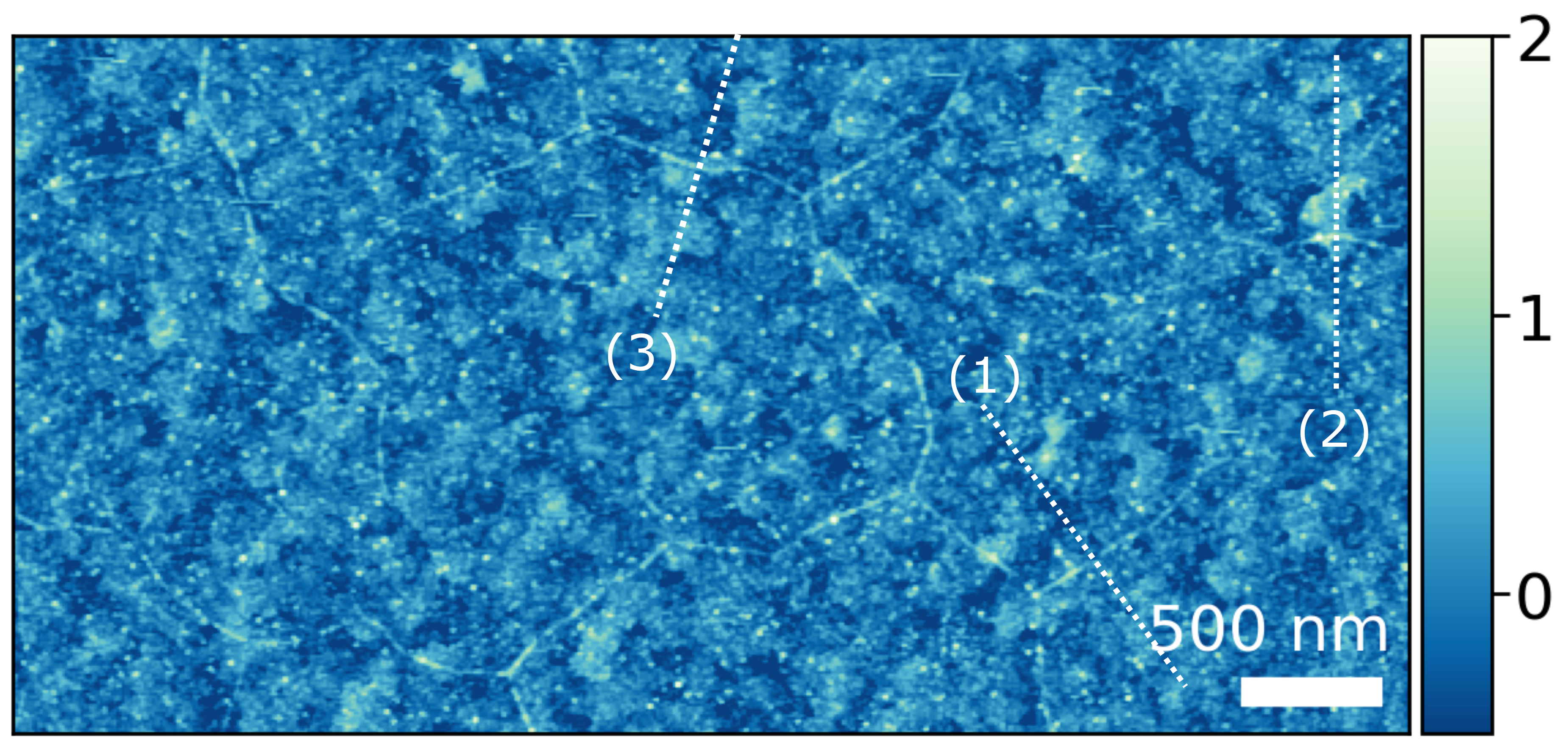}
		\label{LineProfiles}
	} \\
	\subfloat[]{
		\includegraphics[height=4cm]{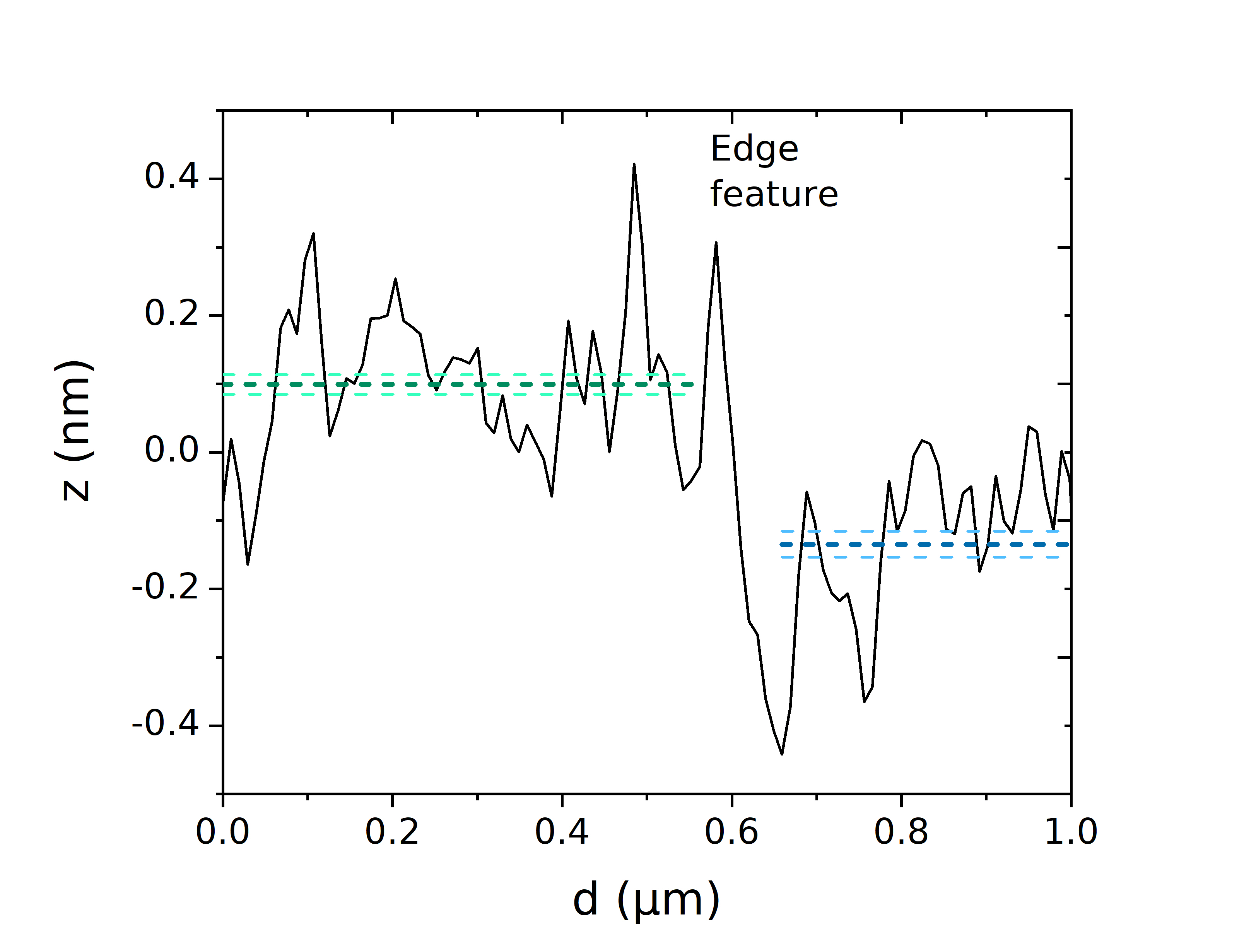}
		\label{Line1}
	}
	\subfloat[]{
		\includegraphics[height=4cm]{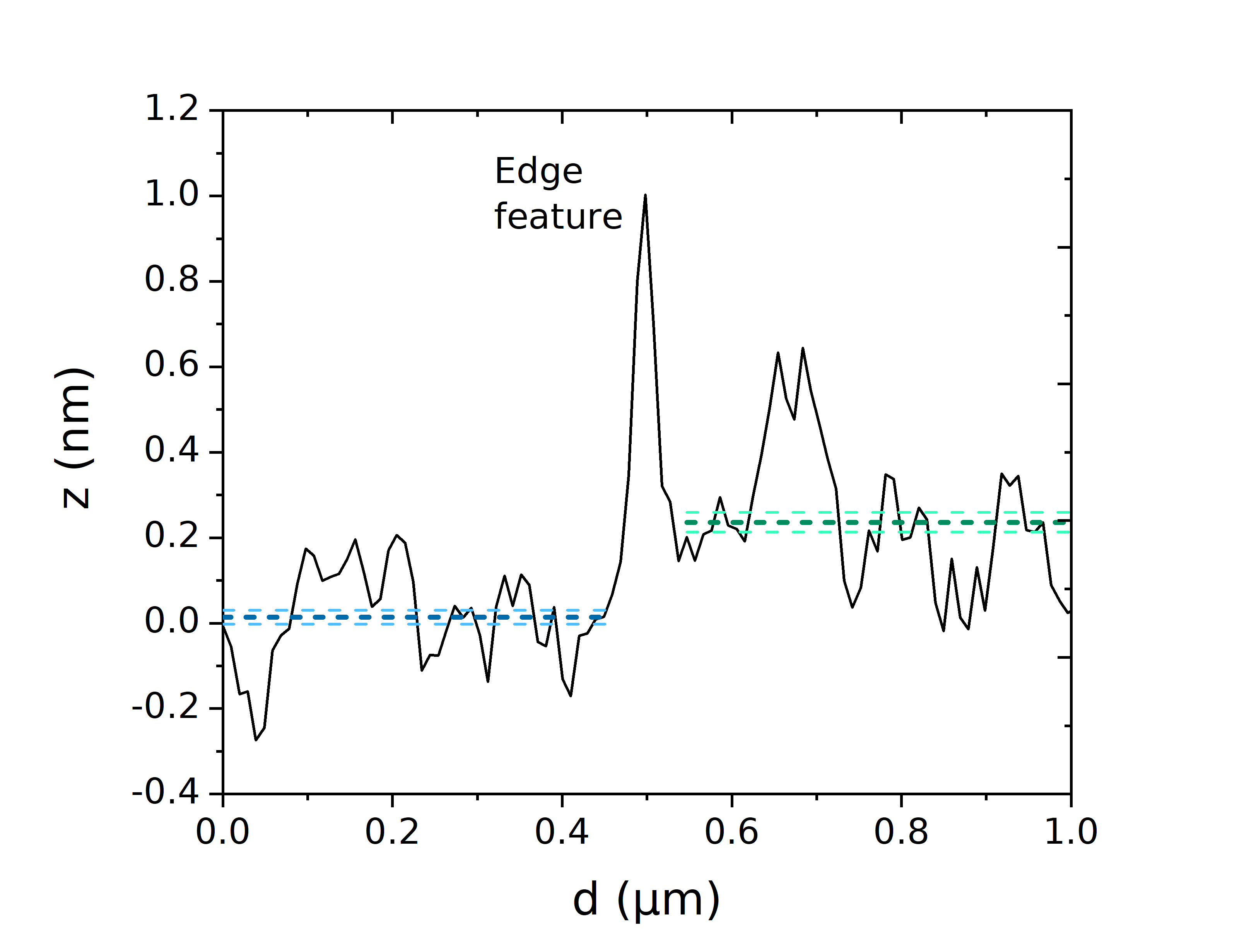}
		\label{Line2}
	}
	\subfloat[]{
		\includegraphics[height=4cm]{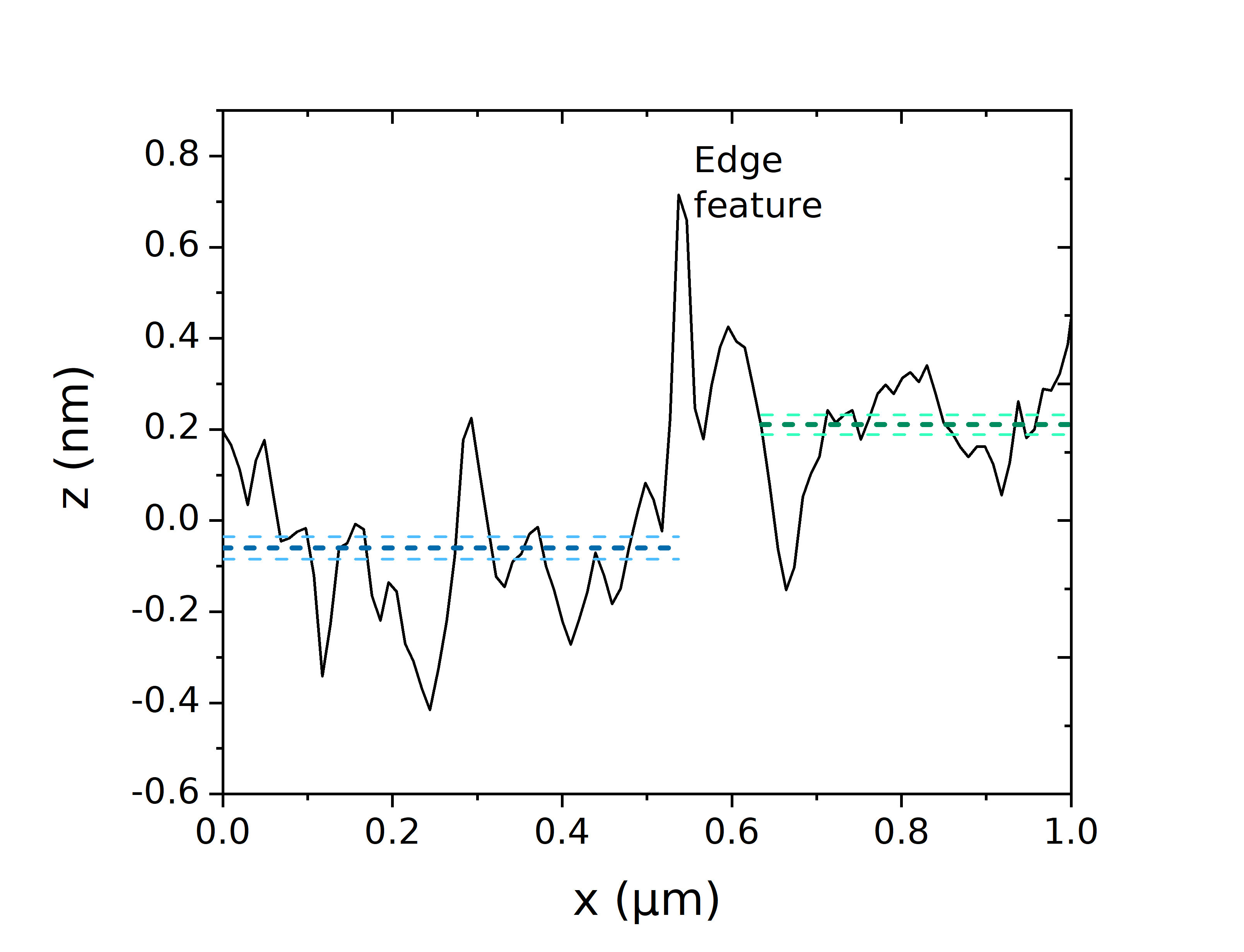}
		\label{Line3}
	}
	\captionsetup{font=footnotesize}
	\caption{Line scans across edge features in AFM maps: \protect\subref{LineProfiles} AFM map of sample surface after HZO deposition directly on graphene, showing the chosen sampling points; extracted line profiles for  \protect\subref{Line1} line 1, \protect\subref{Line2} line 2, \protect\subref{Line3} line 3. The green and blue lines show the mean height (solid) and calculated error (dashed).}
\label{sfig-1_LineScans}
\end{center} 
\end{figure}

Figure \ref{sfig-1_LineScans} shows the AFM map of the surface after ALD deposition of HZO directly on graphene. Features with a higher deposition thickness can be clearly observed. Line profiles were then taken perpendicular to these features, plotted in figures \ref{Line1}-\subref{Line3}. From these, the edge features were first distinguished from comparison with the AFM image and excluded from the analysis. Then, the average sample height either side of the edge features was calculated, along with a standard error. The difference in these heights was found to be: \subref{Line1} $0.234 \pm 0.033 nm$, \subref{Line2} $0.222 \pm 0.040 nm$ and \subref{Line3} $0.270 \pm 0.046 nm$, which agree well with a single monolayer in Pt(111), 0.238 nm \cite{itaya1990nsitu}. Thus, we attribute the edge features to atomic steps in Pt(111). Such steps were observed in scanning tunneling microscopy images on similar Gr/Pt(111)/oxide samples \cite{ajejas2019thermally} (Supporting information). In the same reference, another striking feature observed on the Gr/Pt surface is rotational domains in the Gr layer, which are typically on the order of 50-100 nm, i.e. they are too small to explain the large-scale features observed here ($\sim$1-2 $\mu{}m$). This allows us to rule out enhanced deposition at rotational domain boundaries in graphene.

\clearpage

\section{Characterization of the sample surface after ALD deposition and anneal}
\begin{figure}[h]
\begin{center}
	\subfloat[]{
		\includegraphics[height=4cm]{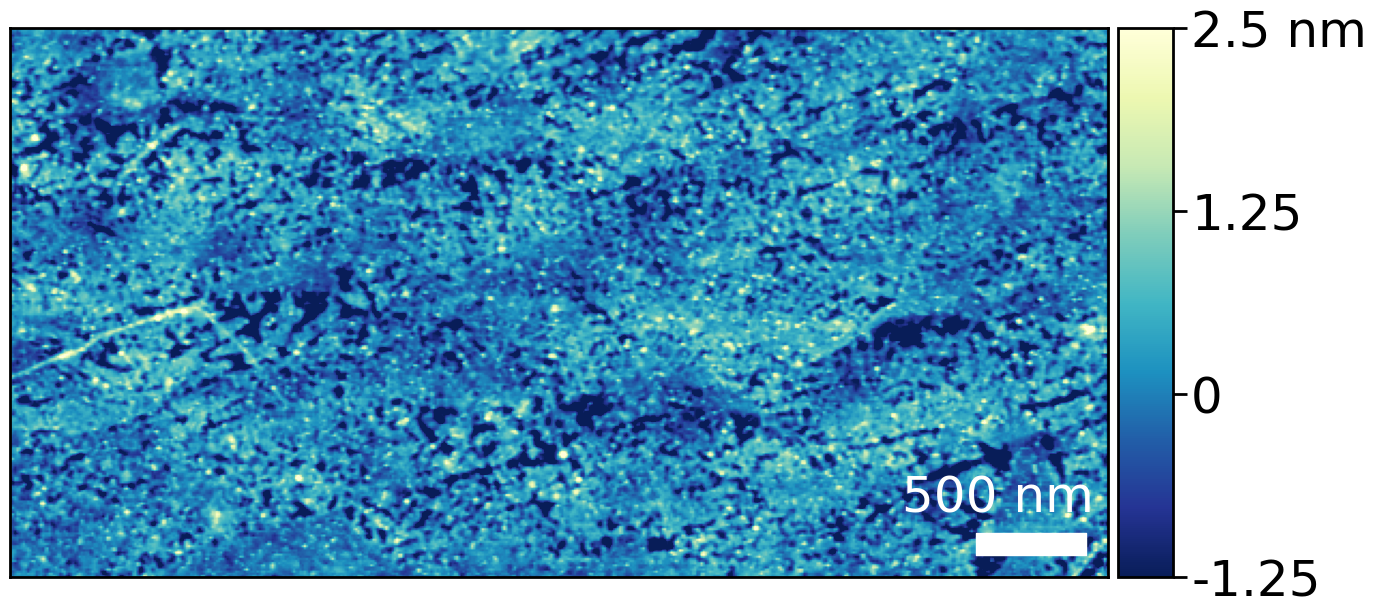}
		\label{afm_AfterGrowth}
	} \\
	\subfloat[]{
		\includegraphics[height=4cm]{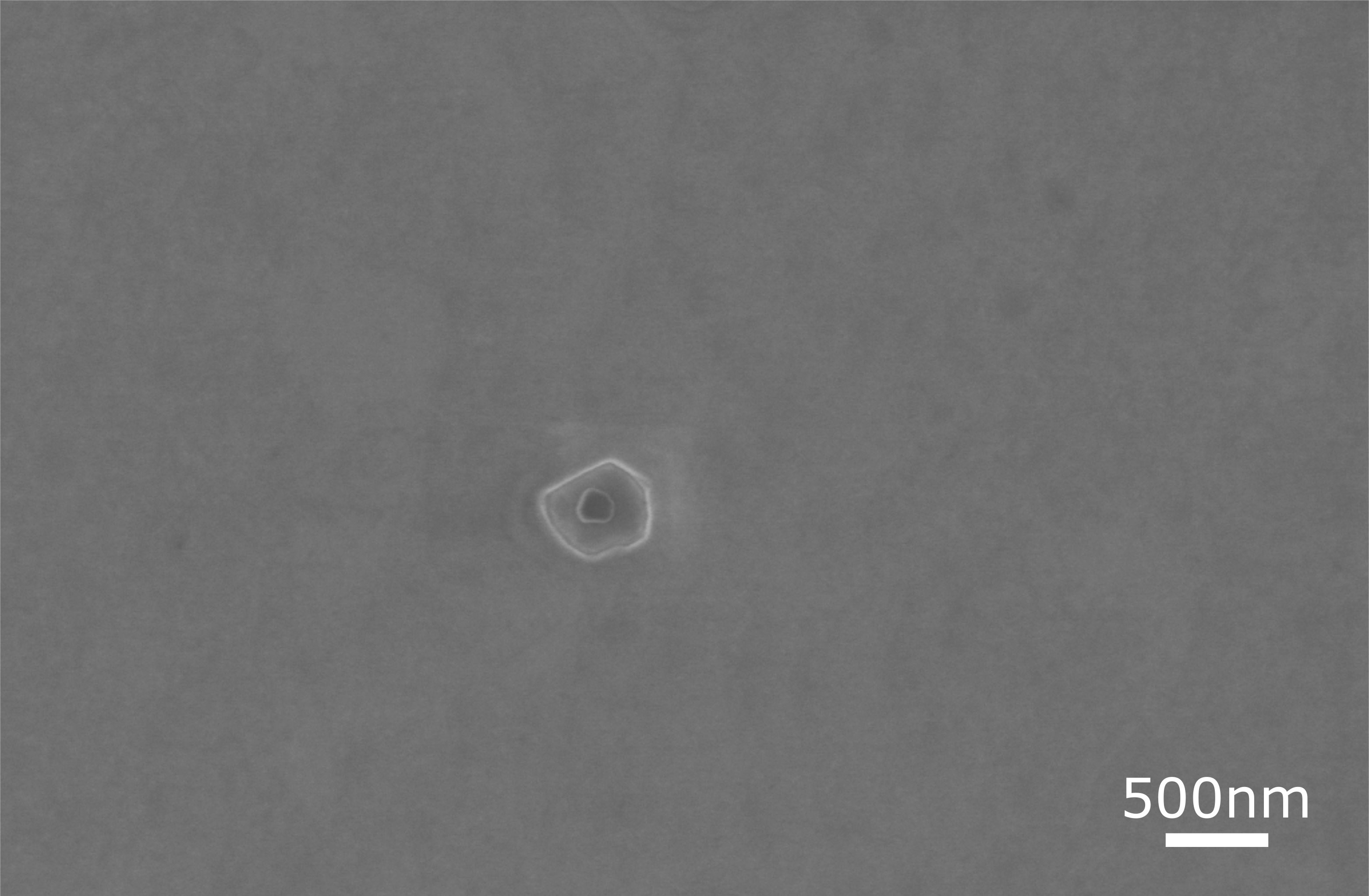}
		\label{sem_AfterGrowth}
	}
	\captionsetup{font=footnotesize}
	\caption{Structural characterisation of final deposited films: \protect\subref{afm_AfterGrowth} AFM and \protect\subref{sem_AfterGrowth} SEM image of \HZO/\AlOx films after crystallisation anneal. In \protect\subref{sem_AfterGrowth} a large defect can be seen at the surface of the HZO film.}
\label{sfig1_final_film}
\end{center} 
\end{figure}

\clearpage
\section{Fitting of GIXRD data} \label{GIXRD_Fit}
\begin{figure}[h]
\begin{center}
	\subfloat[]{
		\includegraphics[height=5cm]{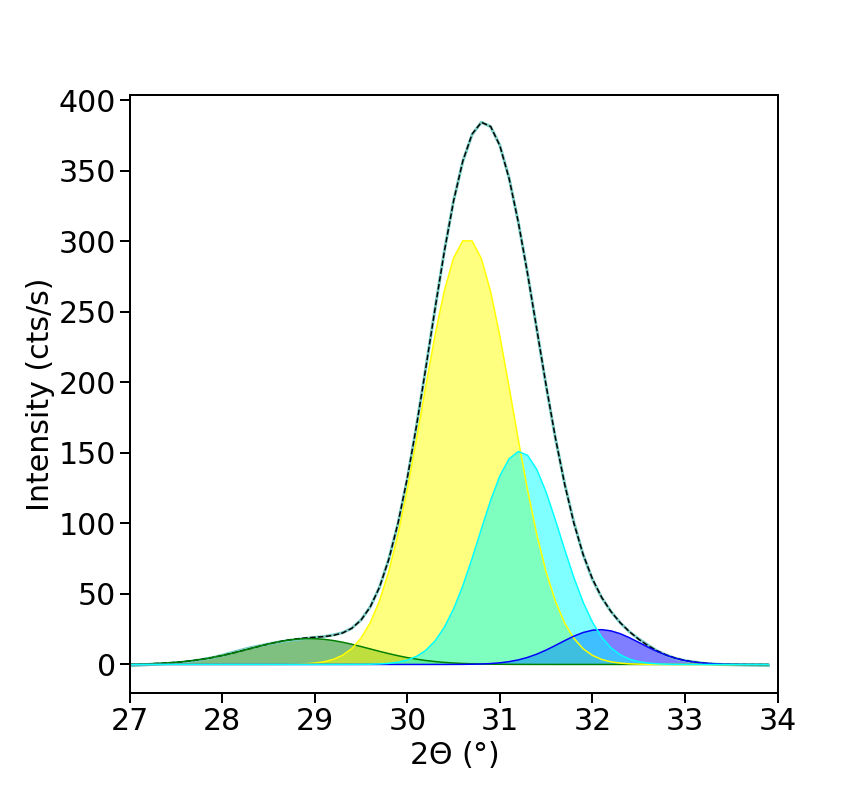}
		\label{GIXRDFit_HZO}
	} 
	\subfloat[]{
		\includegraphics[height=5cm]{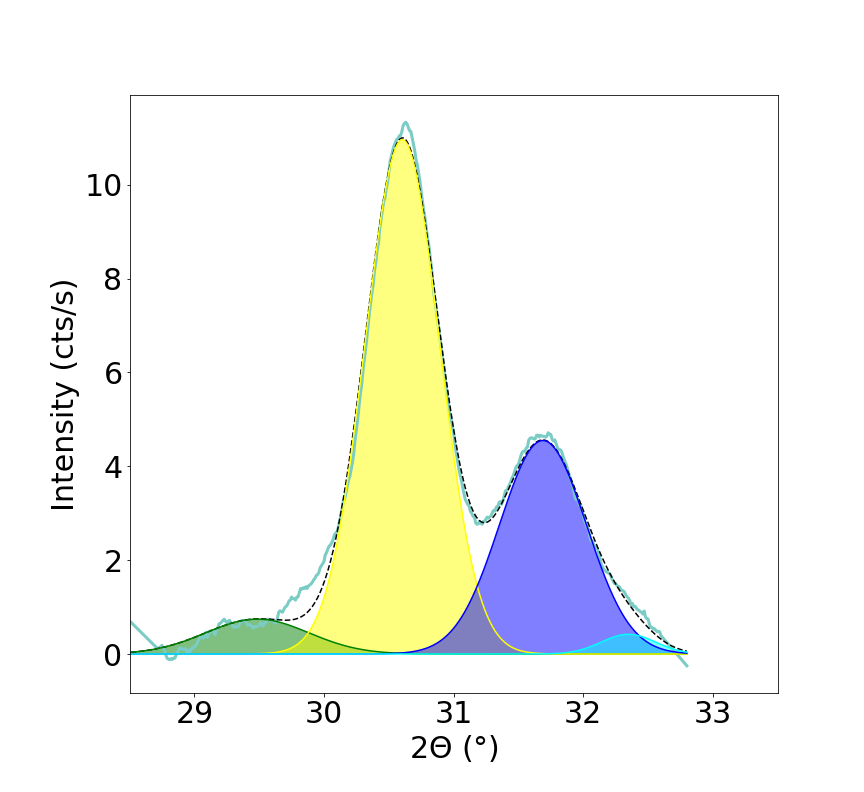}
		\label{GIXRDFit_AlOx}
	} \\
	\subfloat[]{
		\includegraphics[height=5cm]{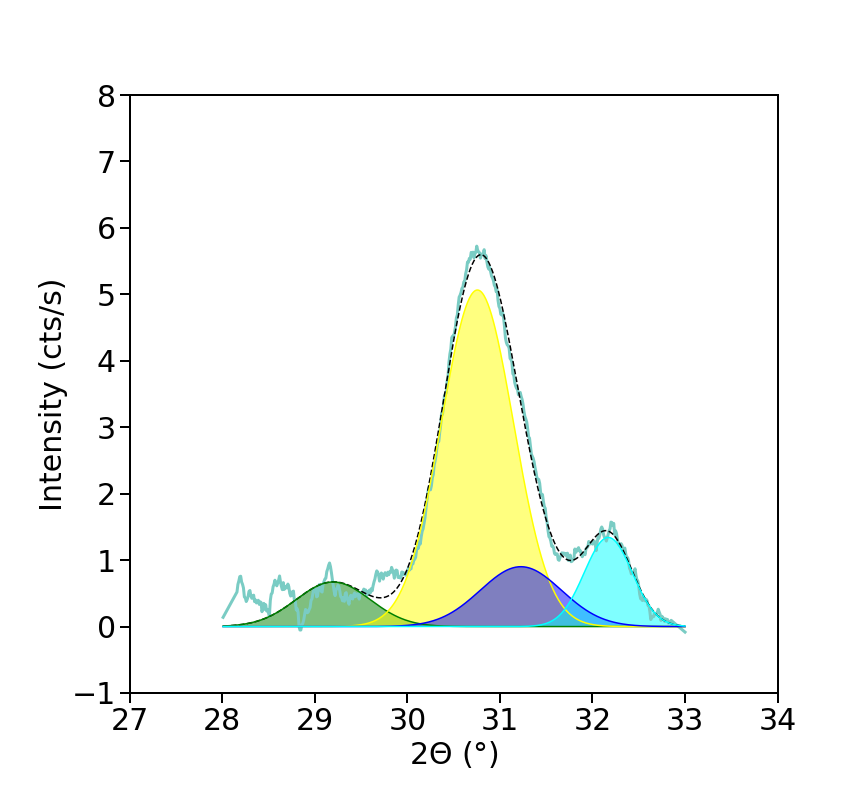}
		\label{GIXRDFit_Ta_Pt}
	} 
	\subfloat[]{
		\includegraphics[height=5cm]{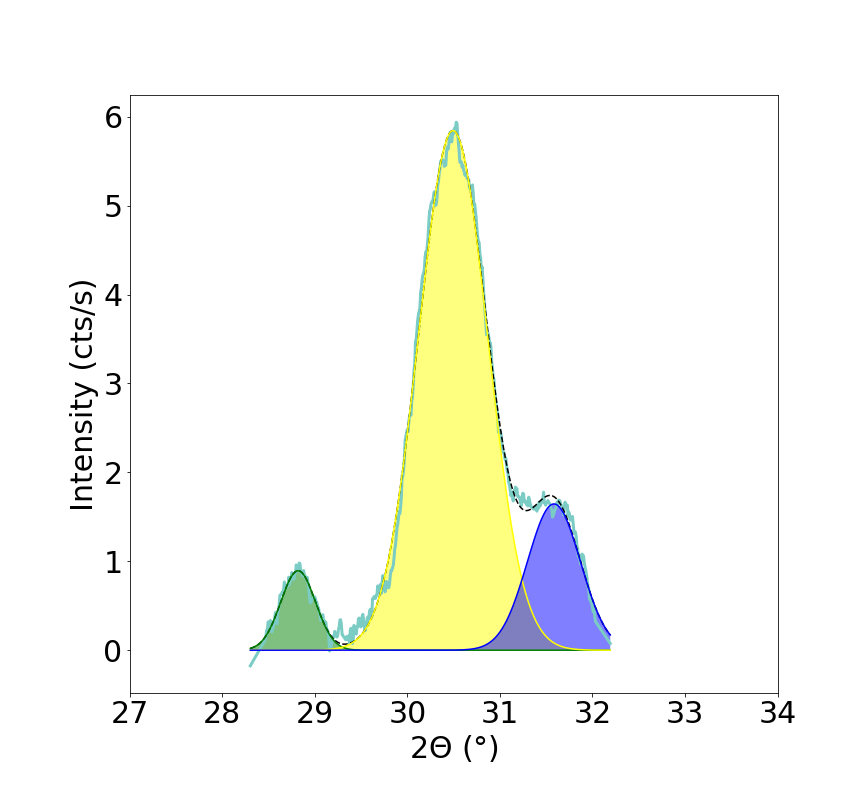}
		\label{GIXRDFit_Ta_Ir}
	} 
	\captionsetup{font=footnotesize}
	\caption{Fitting of peaks in GIXRD: \protect\subref{GIXRDFit_HZO} reference sample of HZO on standard TiN electrodes;\protect\subref{GIXRDFit_AlOx} HZO with Al$_2$O$_3$ as nucleation layer, with Pt as HM; \protect\subref{GIXRDFit_Ta_Pt} HZO with Ta as nucleation layer, with Pt as HM; \protect\subref{GIXRDFit_Ta_Ir} HZO with Ta as nucleation layer, with Ir as HM}
\label{sfig_GIXRD_fitting}
\end{center} 
\end{figure}

The GIXRD data for different films were fitted using a nonlinear least-squares fitting method, assuming Gaussian peaks. The four fitted peaks are the monoclinic (m-)[-111] peak, orthorhombic [111] peak, tetragonal [101] peak and m-[111] peak. The starting peak positions (HfO$_2$ lattice) were taken from Powder Diffraction Cards from the International Centre for Diffraction Data, and the region of interest chosen for fitting is 28-33\degree, where the highest intensity peaks are located. For the fitting procedure, the most well-defined peak (usually the m-phase peak around 31.6°) was fitted first, and the resulting peak width was fixed for all other peaks, assuming broadening is mainly related to grain size and strain. This allowed us to establish reasonable fits for both peak position and peak area, thus allowing an analysis of the relative phase fractions, summarized in Table \ref{Fit_summary}. \\

\begin{table}[]
\begin{center}
\begin{tabular}{|ccccccc|}
\hline
\multicolumn{4}{|c||}{Peak position (2$\theta$)}                                                                                         & \multicolumn{3}{c|}{\textbf{Relative phase fraction (\%)}}      \\ \hline
\multicolumn{1}{|p{0.1\textwidth}|}{m{[}-111{]}} & \multicolumn{1}{p{0.1\textwidth}|}{o{[}111{]}} & \multicolumn{1}{p{0.1\textwidth}|}{t{[}101{]}} & \multicolumn{1}{p{0.1\textwidth}||}{m{[}111{]}} & \multicolumn{1}{p{0.1\textwidth}|}{\textbf{m}}  & \multicolumn{1}{p{0.1\textwidth}|}{\textbf{o}}  & \multicolumn{1}{p{0.1\textwidth}|}{\textbf{t}}  \\ \hline
\multicolumn{7}{|c|}{(a) Reference: HZO on TiN}                                                                                                                                                      \\ \hline
\multicolumn{1}{|c|}{28.82}        & \multicolumn{1}{c|}{30.72}       & \multicolumn{1}{c|}{31.07}       & \multicolumn{1}{c||}{32.06}       & \multicolumn{1}{c|}{\textbf{8}} & \multicolumn{1}{c|}{\textbf{64}} & \multicolumn{1}{c|}{\textbf{28}} \\ \hline
\multicolumn{7}{|c|}{(b) HZO on STO(substrate)/Ta/Pt/Graphene/Al2O3}                                                                                                                                              \\ \hline
\multicolumn{1}{|c|}{28.56}            & \multicolumn{2}{c|}{30.6}           &  \multicolumn{1}{c||}{31.72}           & \multicolumn{1}{c|}{\textbf{34}}   & \multicolumn{2}{c|}{\textbf{66*}}  \\ \hline
\multicolumn{7}{|c|}{(c) HZO on STO(substrate)/Ta/Pt/Co/Graphene/Ta (3 nm)}                                                                                                                                       \\ \hline
\multicolumn{1}{|c|}{29.4}        & \multicolumn{1}{c|}{30.65}       & \multicolumn{1}{c|}{31.15}       & \multicolumn{1}{c||}{32.13}       & \multicolumn{1}{c|}{\textbf{23}} & \multicolumn{1}{c|}{\textbf{49}} & \multicolumn{1}{c|}{\textbf{28}} \\ \hline
\multicolumn{7}{|c|}{(d) HZO on STO(substrate)/Ir/Co/Graphene/Ta (2 nm)}                                                                                                                                       \\ \hline
\multicolumn{1}{|c|}{28.84}        & \multicolumn{1}{c|}{30.29}       & \multicolumn{1}{c|}{30.69}       & \multicolumn{1}{c||}{31.56}       & \multicolumn{1}{c|}{\textbf{26}}  & \multicolumn{1}{c|}{\textbf{37}} & \multicolumn{1}{c|}{\textbf{37}} \\ \hline
\multicolumn{7}{|c|}{\textit{*Mixed o-/t-phase fitted for sample on \AlOx{}}} \\ \hline

\end{tabular}
\label{Fit_summary}
\caption{Summary of peak positions and relative phase fractions extracted from fitting of GIXRD data}
\end{center}
\end{table}

All films deposited on graphene show a reduced orthorhombic phase fraction and thus a lower maximal achievable Pr. This is due to two effects. The first concerns the formation of oxygen vacancies in the HZO. In all cases, the films were annealed with an oxide TE, which may hinder the formation of oxygen vacancies, known to be critical for stabilization of the orthorhombic phase. Nonetheless, the tetragonal phase fraction remains stable, which also requires a large number of oxygen vacancies. One hypothesis is that the HZO in contact with TaOx forms an interfacial tetragonal layer \cite{cheng2022reversible} due to oxygen scavenging from TaOx during annealing. The second effect is the strain state of the stack, which impacts the formation of the orthorhombic phase. Besides peak position, this affects the grain size, visible through a narrowing of the peaks in GIXRD (observed on all samples deposited on Gr). It has been demonstrated computationally that smaller grains favour the formation of the tetragonal phase \cite{materlik2015origin}. \\

\clearpage
\section{Electrical characterization of 20 nm \HZO{} with \AlOx{} crystallization layer intact}
\begin{figure}[!bht]
\begin{center}
	\subfloat[]{
		\includegraphics[width=8cm]{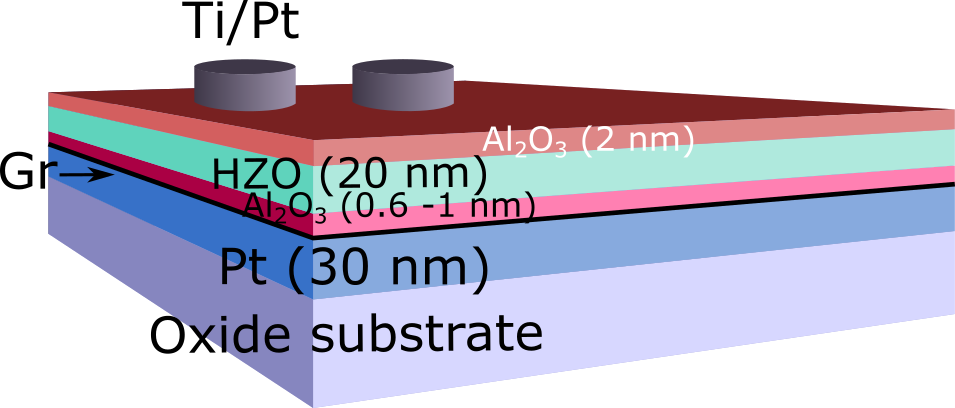}
		\label{AlOx_UnetchedCaps}
	} 
		\subfloat[]{
		\includegraphics[width=8cm]{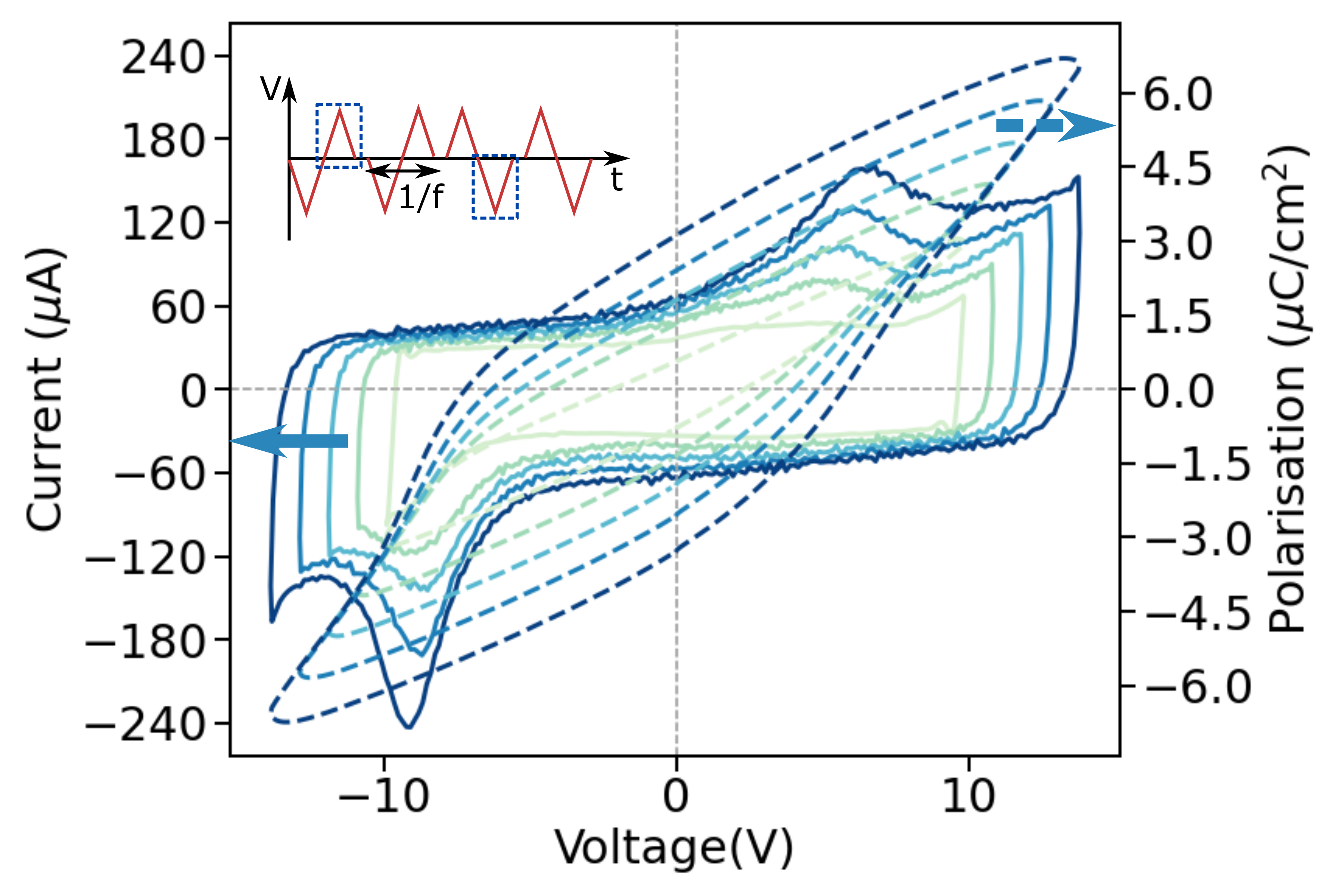}
		\label{AlOx_UnetchedIV}
	}
	\captionsetup{font=footnotesize}
	\caption{Electrical characterization of HZO films: \protect\subref{AlOx_UnetchedCaps} schematic and  \protect\subref{AlOx_UnetchedIV} measured switching characteristics of devices with 20\,nm HZO deposited after 6 cycles \AlOx{} for nucleation, with top \AlOx{} layer left unetched (inset: DHM pulse sequence, with blue dashed lines highlighting the pulses plotted here)}
\label{figSx_Unetched}
\end{center} 
\end{figure}

Figure \ref{AlOx_UnetchedCaps} shows a schematic of the sample stack for HZO on Al$_2$O$_3$ before etching, i.e. where the top Al$_2$O$_3$ layer used for the crystallization anneal is intact. The switching properties are shown in figure \ref{AlOx_UnetchedIV}. The sample demonstrates asymmetric switching properties, with a positive/negative coercive voltage \textit{V$_c^{+/-}$} of +5.7/-7.2 V and remanent polarization \textit{P$_r^{+/-}$} of +3.1 and -3.3 \PrUnits, respectively. The high coercive voltages are related to the \AlOx{} layers \cite{lomenzo2019ferroelectric}. Since the plotted voltage is applied across the whole stack (including \AlOx{}), a certain voltage is dropped across these interlayers, depending on their dielectric constant and thickness, which decreases the voltage drop across the FE layer and thus increases the apparent coercive voltage extracted from I-V and P-V curves. The switching curves are shown for 10-14\,V; below this, the devices didn't switch, and at higher voltages, they broke down.

\clearpage
\section{Electrical characterization of 10 nm \HZO{} deposited on an interlayer of \AlOx{}}
\begin{figure}[h]
\begin{center}
	\subfloat[]{
		\includegraphics[height=5cm]{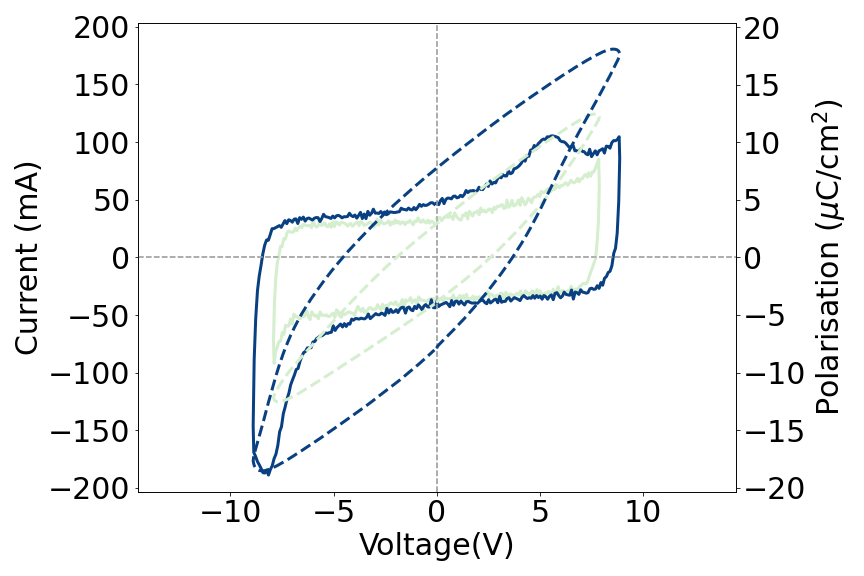}
		\label{Elec_10nmOx}
	} \\
	\captionsetup{font=footnotesize}
	\caption{Electrical measurements on 10 nm HZO on an \AlOx nucleation layer: Polarization-voltage and Current-voltage curves measured at $V_{max}$ = 8-9 V. At 9 V, ferroelectric switching is observed. Above this, the devices break down. The devices are unetched, i.e. the top 2\,nm \AlOx{} layer used for crystallization was left intact.}
\label{sfig2_IV_10nm}
\end{center} 
\end{figure}

\section{HRTEM and EELs analysis of stacks} 
\begin{figure}[!bth]
    \centering
    \subfloat[]{
        \includegraphics[height=5.5cm]{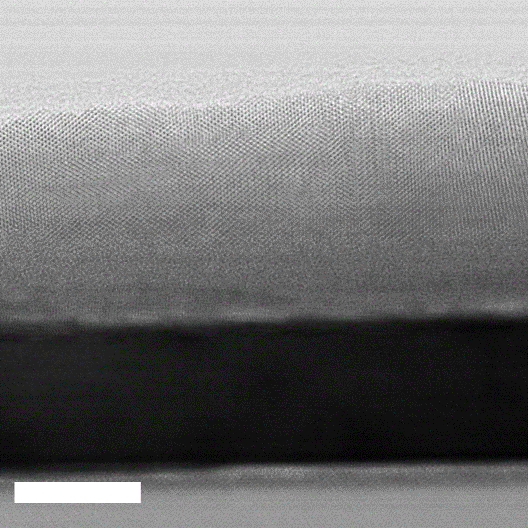}
        \label{HRTEM_full}
    }
    \subfloat[]{
        \includegraphics[height=5.5cm]{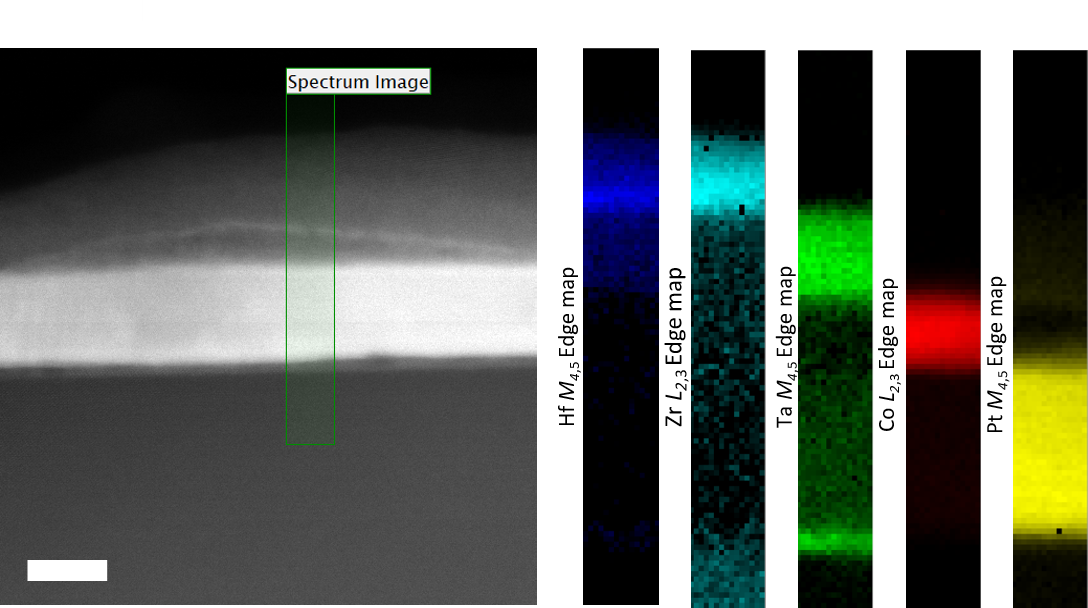}
        \label{EELS}
    }

    \caption{Cross-section STEM-EELS study of deposited films: \protect\subref{HRTEM_full} high resolution annular bright field STEM images of the full stack (HZO deposited on TaOx/Gr/Co/Pt(111)), showing thickness homogeneity and the polycrystalline structure of the HZO film; \protect\subref{EELS} (Left) High angle annular dark field STEM image of the stack. A green rectangle shows the area where an EEL spectrum image was acquired. (Right) EELS maps showing the spatial distribution of the Hf, Zr, Ta, Co ant Pt elemental signals, extracted from the analysis of the absorption edges marked on the panels. The chemical composition mapping is not normalized and is non-quantitative. The scale bars represent 10 nm in all cases.}
   \label{sfig3_TEM}
\end{figure}

\section{Magnetic hystereis measurements on stacks with Ir(111)}
\begin{figure}[!hbt]
    \centering
    \includegraphics[height=6cm]{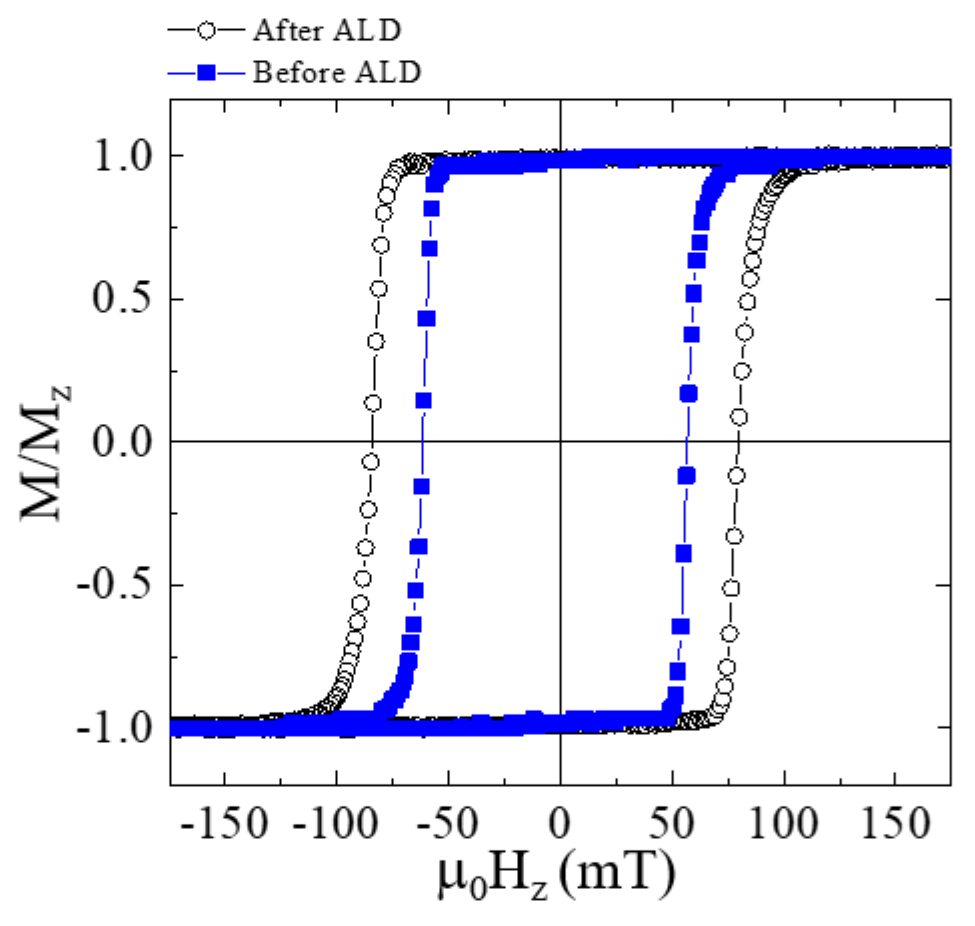}
    \caption{Out-of-plane magnetization component $M_z$ normalized to the saturation value $M_S$ as a function of the applied perpendicular magnetic field $\upmu_0$H$_z$ acquired by magneto-optical (polar) Kerr magnetometry before (blue filled squares) and after (empty black circles) the ALD deposition for 1.5 nm Co layer sandwiched between Gr and Ir(111) buffer. TaOx was used as a nucleation layer in this sample.}
   \label{sfig4_mag_hyst}
\end{figure}

\bibliographystyle{unsrt}
\bibliography{HZO_On_graphene}
